\documentclass[a4paper,11pt]{article}
\usepackage[margin=2.5cm]{geometry}
\usepackage{graphicx}
\usepackage{epsfig}
\usepackage{amsmath}
\usepackage{amssymb}
\usepackage{cite}
\usepackage{float}
\usepackage[font={footnotesize,it}]{caption}

\begin{document}

\begin{titlepage}
\title{p-wave Holographic Superconductors from Born-Infeld Black Holes}
\author{}
\date{
Pankaj Chaturvedi and Gautam Sengupta
\thanks{\noindent E-mail:~ cpankaj@iitk.ac.in , sengupta @iitk.ac.in}
\vskip0.4cm
{\sl Department of Physics, \\
Indian Institute of Technology,\\
Kanpur 208016, \\
India}}
\maketitle
\abstract{
\noindent

We obtain (2+1) dimensional p-wave holographic superconductors from charged Born-Infeld black holes in the presence of massive charged vector fields in a bulk $AdS_4$ Einstein-Born-Infeld theory through the $AdS_4$-$CFT_3$ correspondence. Below a certain critical transition temperature the charged black hole develops vector hair that corresponds to charged vector condensate in the strongly coupled (2+1) dimensional boundary field theory that breaks both the $U(1)$ symmetry as well as the rotational invariance. The holographic free energy is computed for the boundary field theory which shows that the vector order parameter exhibits a rich phase structure involving zeroth order, first order, second order and retrograde phase transitions for different values of the backreaction and the Born-Infeld parameters.  We numerically compute the ac conductivity for the p-wave superconducting phase of the strongly coupled (2+1) dimensional boundary field theory which also depends on the relative values of the parameters in the theory.}

\end{titlepage}
\section{Introduction}

The AdS-CFT or the gauge/gravity correspondence is one of the most significant developments in fundamental physics in the last one decade. This holographically relates a weakly coupled theory of gravity in a bulk  AdS space time to a strongly coupled gauge theory at its conformal fixed point on the conformal boundary in a large N limit and vice versa \cite{Maldacena:1997re,Gubser:1998bc,Witten:1998qj,Aharony:1999ti}. Later studies showed that the strongly coupled conformal field theory on the boundary may be described at a finite temperature through the introduction of a black hole in the bulk AdS space time \cite {Witten:1998zw}. The Hawking temperature of the bulk AdS black hole was then considered to be the temperature of the boundary gauge theory.  In particular the Hawking-Page phase transition for the black hole in the bulk AdS space-time corresponds to the confinement-deconfinement phase transition in the boundary gauge theory \cite {Witten:1998zw}.  In recent years such strongly coupled boundary field theories at a finite temperature and a finite charge density could be realized through this correspondence from charged black holes in the bulk AdS space-time coupled to gauge fields and other matter fields. This makes the correspondence an extremely powerful tool for describing strongly coupled condensed matter systems,  which are typically at finite temperatures and finite density.   In this context Gubser \cite{Gubser:2008px} showed that a charged Reissner Nordstrom-AdS (RN-AdS) black hole in presence of a charged scalar field was unstable to the formation of scalar hair below a certain critical temperature. Through the AdS-CFT correspondence Hartnoll et al. \cite{Hartnoll:2008vx} demonstrated that this bulk instability corresponds to a superconducting phase transition in the boundary field theory. Such strongly coupled superconducting phases of the boundary field theory are termed {\it holographic superconductors}. The construction described above relates to a s-wave holographic superconductor and typically involves an Abelian Higgs model with a bulk complex scalar field that is charged under the Maxwell field \cite{Hartnoll:2008kx,Herzog:2009xv,Horowitz:2010gk,Herzog:2010vz,Sachdev:2010ch}. From a phenomenological perspective, the onset of superconductivity in the boundary field theory causes a charged operator to acquire a vacuum expectation value which spontaneously breaks the global $U(1)$ invariance. It is to be noted that the dual boundary field theory is translationally invariant which leads to the divergence of the Drude peak at zero frequency and a gap formation in the real part of the electrical conductivity which is characteristic of such superconducting phase transitions \cite{Roberts:2008ns,Horowitz:2009ij}. The physics of this transition could be described in the probe approximation where the backreaction of matter fields on the spacetime metric is neglected \cite{Hartnoll:2008vx}. It was later shown that this backreaction may be systematically accounted for and leads to a lower transition temperature thereby making the condensation harder\cite {Hartnoll:2008kx}.

Naturally such a model of superconductivity involving the AdS-CFT correspondence is a possible candidate for the description of high temperature superconductors which are typically strongly coupled. It is also known that such high temperature superconductors usually involve p, d or higher wave superconductors. The construction for holographic s-wave superconductors may be generalized to describe  d-wave superconductors through the condensation of a charged massive spin two field in the bulk \cite{Kim:2013oba,Benini:2010pr,Chen:2010mk}. Consequently p-wave holographic superconductors could be realized through the condensation of a charged vector field in the bulk that corresponds to a dual vector order parameter in the boundary field theory. It was shown in \cite{Gubser:2008wv} that such a  holographic p-wave superconductor may also be described through a bulk $SU(2)$ Yang-Mills gauge field where one of the gauge degrees of freedom for the $SU(2)$ gauge field is  dual to the vector order parameter in the boundary field theory. Further studies based on this model involving condensate formation, transport and spectral properties in diverse dimensions both in the probe limit and including the back reaction may be found in \cite{Ammon:2009xh,Zeng:2010fs,Cai:2010zm,Aprile:2010ge,Zayas:2011dw,Momeni:2012ab,Roychowdhury:2013aua,Cai:2013oma}. Alternatively such models of p wave holographic superconductors may be realized through the condensation of a 2-form field in the bulk \cite{Cai:2013pda} or through the condensation of a massive complex vector field charged under the Maxwell field  \cite{Cai:2013aca}. In \cite{Ammon:2009xh} it was shown that for the model of p-wave holographic superconductors, the phase transition in the boundary field theory corresponding to the formation of vector hair for the bulk charged black hole changes from second order to first order for  large backreaction of the matter fields. Whereas in \cite{Cai:2013aca}, it was observed that depending on the mass of the complex vector field and the strength of  the backreaction, the condensate formation may occur through a first order, second order, zeroth order \cite{Maslov:2004ap} or a retrograde phase transition \cite{Narayanan1994135}. Such a varied phase structure may also be observed for the s-wave holographic superconductors as has been shown in the articles \cite{Horowitz:2010jq,Peng:2011gh,Cai:2012es,Horowitz:2008bn}.

It is to be noted that all of the above mentioned gravity duals for the holographic superconductors have been studied in the framework of linear Maxwell electrodynamics related to the Einstein-Maxwell gravity in the bulk. It is naturally interesting to investigate the possibility of  describing such  holographic superconductor in a non linear scenario. In this context the Born-Infeld electrodynamics \cite{Born:1934trs} is one of the most important nonlinear electromagnetic theories that was proposed to avoid the infinite self energy arising in the Maxwell theory. Moreover, the Born-Infeld action naturally possesses electric-magnetic duality invariance \cite{Gibbons1995185} which makes it suitable for describing gauge fields arising from open strings on D-branes  \cite{Fradkin1985123}. In \cite{Dey:2004yt,Cai:2004eh} asymptotically Anti-deSitter (and deSitter) black hole solutions were obtained for the Einstein Born-Infeld theory with a cosmological constant. The condensate formation and the transport properties for holographic s-wave superconductors in the framework of  the nonlinear Born-Infeld electrodynamics have been studied extensively in \cite{Jing201068,Gangopadhyay:2013qza,Jing:2011sv,PhysRevD.83.066010,Gangopadhyay:2012sv,Liu:2012hc,PhysRevD.86.106009,Zhao201398,yao2013sv}. Specifically, in \cite{Jing201068} it was shown that in the probe limit, for large values of the Born-Infeld parameter,  the condensate formation is hindered while suppressing the gap in the real part of the ac conductivity.

In this article we propose to construct a model of  p-wave holographic superconductors in the framework of  an Einstein-Born-Infeld theory in the bulk AdS space time with backreaction. In this context we consider a charged Born-Infeld black hole in the presence of a massive complex vector field that is charged under the Maxwell gauge field in a bulk $AdS_4$ Einstein-Born-Infeld gravity. Through the AdS-CFT correspondence such a bulk theory would be dual to a strongly coupled (2+1) dimensional boundary field theory involving a charged vector operator with a global $U(1)$ symmetry. It is also possible  to include a non-minimal coupling \cite{Cai:2013aca} between the vector field and the gauge field in the bulk Einstein Born-Infeld theory that represents the magnetic moment of the complex vector field and plays a crucial role for the condensate formation induced by an external magnetic field. As our motivation here is to study the effect of the Born-Infeld parameter on the p-wave holographic superconductors we have neglected this coupling in our analysis. We show that below a certain critical temperature the dual vector operator in the (2+1) dimensional boundary field theory acquires a vacuum expectation value that breaks the $U(1)$ symmetry as well as the rotational symmetry at the boundary spontaneously that essentially describes an anisotropic condensate required for a p-wave superconductor. Naturally such a condensate is a consequence of the formation of vector hair for the charged Born-Infeld black hole in the $AdS_4$ bulk. Our construction involves  three independent parameters namely, the mass $m$ of the vector field which is related to the dimension of the dual vector operator, the back reaction parameter $\kappa$ and the Born-Infeld parameter $\gamma$ all of which control the phase structure of the (2+1)-dimensional boundary field theory. Depending on the values of the parameters 
$m$, $ \kappa$ and $\gamma$ we realize a rich and varied phase structure for the boundary field theory in our construction. We further observe that for fixed values of  $m$  and $\kappa$ it is possible to tune the parameter  $\gamma$ and obtain a change in the order of the superconducting phase transition of the boundary field theory. This is in contrast to \cite{Cai:2013aca} as our model whilst reproducing the usual phase structure of such p-wave holographic superconductors also exhibits a novel phase behaviour that arises from the inclusion of the nonlinear Born-Infeld term in the bulk theory.

We observe in our construction that the phase structure of the boundary theory changes radically as we go from large values of  $m^2$ to small values of $m^2$. For large $m^2$ it is observed that at a small but fixed value of the Born-Infeld parameter $\gamma$, the superconducting phase transition changes from second order to first order for a critical value of the  backreaction parameter $\kappa$.  Whereas for small $m^2$, we observe that there is no condensate formation below a certain critical transition temperature for all values of  the parameters $\kappa$ and $\gamma$. Furthermore at small but fixed values of $\gamma$ a small backreaction results in a second order phase transition below a critical temperature. This changes to a zeroth order phase transition at some lower temperature where the free energy changes discretely \cite{Maslov:2004ap}. For moderate values of the backreaction parameter the superconducting phase transition is of first order and there is no condensate formation at a certain temperature below the transition temperature. In contrast for large values of the backreaction parameter we observe an interesting {\it retrograde phase transition} \cite{Narayanan1994135}, for which the condensate formation occurs only for temperatures higher than the transition temperature. In general we see that  for small $m^2$ it is possible to obtain bounds on  the values of $\gamma$ which indicates that the order of the phase transition may be changed through tuning the Born-Infeld parameter $\gamma$ for fixed values of the backreaction parameter $\kappa$. Furthermore it is observed that for all values of the mass $m^2$ the transition temperatures corresponding to different orders of phase transitions decrease with increasing values of $\kappa$ and $\gamma$. 

In contrast to \cite{Cai:2013aca} in a linear Maxwell scenario, we have further computed transport properties such as the ac conductivity of the strongly coupled 2+1 dimensional field boundary theory in a linear response framework. This is the first such computation in the literature for p-wave holographic superconductors in a non linear Born-Infeld setting. The ac conductivity for the different phases of the p-wave holographic superconductors arising in our construction exhibits a dependence on all the three parameters $m$, $ \kappa$ and $\gamma$. Remarkably in the case of the retrograde phase transition described above, we observe the formation of a gap in the real part of the ac conductivity at temperatures above the critical temperature. For the rest of the cases the ac conductivity exhibits a behavior similar to the case of the p-wave holographic superconductors obtained in \cite{Gubser:2008wv} from the condensation of the vector part of an $SU(2)$ gauge field in the $AdS_4$ bulk.

The article is organized as follows, in Section 2 we introduce the bulk gravitational theory and obtain the corresponding equations of motion. In this section we also describe the charged Born-Infeld black hole solutions with vector hair in the bulk   $AdS_4$ space-time which corresponds to the p -wave superconducting phase of the boundary field theory. In section 3 we compute the free energy and the dual stress energy tensor for the 2+1 dimensional boundary field theory using the bulk-boundary correspondence and the holographic dictionary. Section 4 is devoted to calculating the condensates corresponding to the different types of phase transitions for different values of the parameters $m^2,\kappa$ and $\gamma$. In section 5 we describe the ac conductivity for the superconducting phase of the boundary theory at various values of the parameters $m^2,\kappa$ and $\gamma$. In Section 6 we present a summary of our results and discussions.

\section{The bulk $AdS_4$ Einstein-Born-Infeld theory}
In order to realize a vector condensate in the Einstein-Born-Infeld 
theory we consider a bulk gravitational action with a complex vector field \cite{Cai:2013aca} that is charged under the bulk Maxwell field associated with the Born-Infeld Lagrangian. The form of the bulk gravitational action may be written down as,
\begin{eqnarray} \label{action}
S &=& \frac{1}{2 \kappa^2}\int dx^{4} \sqrt{- g}\left({\cal L}_G \right)+\int dx^{4}\sqrt{- g}\left({\cal L}_M \right)\nonumber\\
{\cal L}_G &=&{\cal R}+\frac{6}{L^2} \nonumber\\ 
{\cal L}_M &=& \frac{1}{\gamma}\left(1-\sqrt{1+\frac{\gamma}{2}F}\right)-\frac{1}{2}\rho^{\dagger}_{\mu\nu}\rho^{\mu\nu}-m^2\rho^{\dagger}_{\mu}\rho^{\mu} -i \eta q\left(\rho^{\dagger}_{\mu}\rho_{\nu}F^{\mu\nu}\right)
\end{eqnarray} 
with L as the AdS radius and $\kappa^2= 8 \pi G$ is related to the gravitational constant in the bulk. Here $\rho_{\mu}$ is the complex vector field charged under the bulk Maxwell field $A_{\mu}$ with $F=F_{\mu\nu}F^{\mu\nu}$. The constants $q$ and $m$ correspond respectively to the charge and the mass of the vector field $\rho_\mu$. The Maxwell field strength reads as $F_{\mu\nu}=\nabla_{\mu}A_{\nu}-\nabla_{\nu}A_{\mu}$ and $\rho_{\mu\nu}$ in (\ref{action}) may be defined as $\rho_{\mu\nu}=D_{\mu}\rho_{\nu}-D_{\nu}\rho_{\mu}$ with the covariant derivative $D_{\mu}=\nabla_{\mu}-i q A_{\mu}$. The last term in the expression for the matter Lagrangian represents the non-minimal coupling between the vector field $\rho_{\mu}$ and the gauge field $A_{\mu}$. This characterizes the magnetic moment of the vector field $\rho_{\mu}$ as suggested by the authors in \cite{Cai:2013pda} where it was shown that the term plays an important role for the case of an applied magnetic field. In our study we will neglect the effect of the interaction term on the boundary field theory 
and simply consider only the effect of the nonlinear Born-Infeld term. For this purpose we will set the interaction parameter $\eta$ to zero at the level of the ansatz for solving the equations of motion.

In our model we consider the backreaction of the bulk fields on the background metric 
that describes a charged Born-Infeld black hole in the $AdS_4$ bulk.  We rescale the bulk fields $\rho_{\mu}$, $A_{\mu}$ and the Born-Infeld coupling parameter $\gamma$ as $\rho_{\mu}/q$, $A_{\mu}/q$ and $q^2\gamma$ in order to put the factor $\frac{1}{q^2}$ as the backreaction parameter for the matter fields. In general the probe limit is defined as $\kappa^2/q^2\rightarrow 0$. In the literature there exist two methods to include the backreaction of matter fields on the metric. The first method is to consider $\kappa^2=1$ and choose a finite value of $q^2$ as described in \cite{Hartnoll:2008kx}. The second alternative is to fix $q^2=1$ and consider finite values of the parameter $\kappa^2$. In our analysis we will use the second approach to fix the backreaction parameter to be $\kappa^2$. We find the equations of the motion for the bulk matter fields through the variation of the action (\ref{action}) as
\begin{eqnarray}
\nabla^{\nu} \left(\frac{F^{\nu\mu}}{\sqrt{1+\frac{\gamma}{2}F}}\right) &=& i (\rho_{\nu}\rho^{\dagger\nu\mu}-\rho_{\nu}^{\dagger}\rho^{\nu\mu}) - i  \eta \nabla_{\nu}(\rho^{\nu}\rho^{\dagger\mu}-\rho^{\dagger\nu}\rho^{\mu}),\label{MaxEom}\\
D_{\nu}\rho^{\nu\mu} &=& m^{2}\rho^{\mu} + i \eta \rho_{\nu}F^{\nu\mu},\label{RhoEom}
\end{eqnarray}
similarly the gravitational field equations may be expressed as,
\begin{eqnarray}
 {\cal R}_{\mu\nu}-\frac{1}{2}{\cal R} g_{\mu\nu}-\frac{3}{L^2} g_{\mu\nu} &=& \frac{\kappa^2}{2}{\cal L}_M g_{\mu\nu}+\frac{\kappa^2}{2} \left(\frac{F_{\mu\lambda}F^{~\lambda}_{\nu}}{\sqrt{1+\frac{\gamma}{2}F}}\right)+\frac{\kappa^2}{2}[(\rho^{\dagger}_{\mu\lambda}\rho_{\nu}^{~\lambda}+m^2\rho^{\dagger}_{\mu}\rho_{\nu} \nonumber \\
 &-& i \eta(\rho^{\dagger}_{\mu}\rho_{\lambda}-\rho_{\mu}\rho^{\dagger}_{\lambda})F_{\nu}^{~\lambda})+\mu\leftrightarrow\nu],\label{EinsEom}
\end{eqnarray}
where the right hand side of the eq.(\ref{EinsEom}) represents the stress energy tensor $T_{\mu\nu}$. According to the gauge/gravity duality we have a bulk theory of gravity that is holographically  dual to a gauge theory on the conformal boundary of the AdS space. In \cite{Gubser:2008px,Hartnoll:2008vx} it was observed that the charged black hole with scalar hair corresponds  to the superconducting phase of the boundary field theory. More precisely as  the temperature of the charged black hole in the bulk reaches a critical temperature the background develops an instability leading to the formation of scalar hair. This corresponds to a superconducting phase transition in the boundary field theory and leads to a charged scalar operator acquiring a vacuum expectation value.
Our construction described by the action (\ref{action}) admits  a vector field ($\rho_{\mu}$) which is charged under the bulk $U(1)$ gauge field $A_{\mu}$. Following the AdS/CFT dictionary this bulk vector field corresponds to a charged vector operator in the strongly coupled dual boundary field theory. A non zero vacuum expectation value of  this dual charged vector operator will break the $U(1)$ symmetry at the boundary along with the rotational invariance which would describe an anisotropic condensate necessary for the realization of a p-wave superconducting phase.  

\subsection{Equations of motion and Boundary conditions}
In order to realize charged black hole solutions with vector hair in our model we consider the following ansatz for the metric and the bulk fields, 
\begin{eqnarray}
ds^2 &=& -f(r)e^{-\chi(r)}dt^2+\frac{dr^2}{f(r)}+r^2 h(r)dx^2+r^2\frac{dy^2}{h(r)},\label{metric}\\
\rho_{\mu}dx^{\mu} &=& \rho_{x}(r)dx,~~A_{\mu}dx^{\mu}=\phi(r)dt.\label{bulkansatz}
\end{eqnarray}
In the metric ansatz we have introduced a function $h(r)$ in the $xx$ and the $yy$ component of the metric which accounts for the anisotropy induced by the vector field $\rho_{\mu}$. We may also account for specific choice of the ansatz as follows, the non zero $x$ component of the vector field $\rho_{\mu}$ denoted by $\rho_x$ correspond to the expectation value $\left\langle J_x\right\rangle$ of a dual vector operator $J_x$ of the boundary theory. Thus as described above a non zero vacuum expectation value of $J_x$ will break both the $U(1)$ gauge symmetry and the rotational invariance in $x-y$ plane with  $\left\langle J_x\right\rangle$  picking the $x$ direction as special. The Hawking temperature $T_h $ of the black hole is given as
\begin{equation}
T_h=\frac{f'(r_h) e^{-\chi(r_h)/2}}{4 \pi},\label{temp}
\end{equation}
where the horizon $r_h$ is determined by $f(r_h)=0$. As mentioned earlier we set the interaction parameter $\eta$ to zero in the equations (\ref{MaxEom},\ref{RhoEom},\ref{EinsEom}). Now implementing the ansatz (\ref{metric}) and (\ref{bulkansatz}), we may write down the independent bulk equations of motion as,
\small
\begin{eqnarray}
\rho _x''+\rho _x'\left(\frac{f'}{f}-\frac{h'}{h}-\frac{\chi '}{2}\right)-\rho _x \left(\frac{m^2}{f}-\frac{e^{\chi} \phi^2}{f^2}\right)&=& 0,\label{eom1} \\
\phi''+\phi'\left(\frac{2}{r}+\frac{\chi'}{2}\right)-\frac{2\rho_x^{2}\phi\left(1-e^{\chi} \gamma  \phi'^{2}\right)^{3/2}}{r^2 f h}-\frac{2 e^{\chi} \gamma  \phi'^{3}}{r}&=& 0,\label{eom2}\\
\chi '+\frac{r h'^2}{2 h^2}+\kappa^2\left(r \psi '^2+\frac{\rho _x'^2}{r h}+\frac{e^{\chi} \rho_{x}^2 \phi^2}{r f^2 h}\right)&=& 0,\label{eom3}\\
h''+h'\left(\frac{f'}{f}-\frac{h'}{h}-\frac{\chi'}{2}+\frac{2}{r}\right)+\kappa^2 \left(\frac{\rho _x'^2}{r^2}+\frac{m^2 \rho _x^2}{r^2 f}-\frac{e^{\chi} \phi^2 \rho _x^2}{r^2 f^2}\right) &=& 0,\label{eom4}\\
\frac{f'}{f}+\frac{r h'^2}{4 h^2}+\kappa^2\left(-\frac{r}{2 \gamma }+\frac{r}{2 \gamma  \sqrt{1-e^{\chi} \gamma  \phi '^2}}+\frac{m^2\rho_x^{2}}{2 r h}+\frac{e^{\chi}\rho_x^{2}\phi^2}{2 r f h}+\frac{f \rho_x '^{2}}{2 r h}\right)-\frac{3 r}{f}+\frac{1}{r}&=& 0,\label{eom5}
\end{eqnarray} 
\normalsize
where the prime denotes derivative with respect to the coordinate $r$. The above set of equations of motion admit an analytic solution for  $\rho_{\mu}=0$ . The solution thus obtained correspond to the normal phase of the boundary theory that describes a charged Born-Infeld $AdS_4$ black hole \cite{Dey:2004yt,Cai:2004eh} in the bulk with the metric
\begin{eqnarray}
ds^2 &=& -f(r)dt^2+\frac{dr^2}{f(r)}+r^2(dx^2+dy^2),\nonumber\\
\phi(r)&=&U+\frac{\mu}{r}~{}_2F_{1}\left[\frac{1}{4},\frac{1}{2},\frac{5}{4},-\frac{4\gamma\mu^2}{r^4}\right],\nonumber\\
f(r)&=&-\frac{M^2}{r}+\left[\frac{\kappa^2}{6\gamma}+\frac{1}{L^2}\right]r^2-\frac{\kappa^2}{6\gamma}\sqrt{r^4+4\gamma\mu^2}+\frac{4\kappa^2\mu^2}{3r^2}~{}_2F_{1}\left[\frac{1}{4},\frac{1}{2},\frac{5}{4},-\frac{4\gamma\mu^2}{r^4}\right],\label{nrmlphase}
\end{eqnarray}
where $\mu$ and $U$ are integration constants with $M$ being the ADM mass of the black hole. Here the symbol, ${}_2F_{1}[a,b,c,z]$ represents a hypergeometric function. The constant $U$ may be determined by the constraint $\phi(r_h)=0$, at the horizon as
\begin{equation}
U=-\frac{\mu}{r_h}~{}_2F_{1}\left[\frac{1}{4},\frac{1}{2},\frac{5}{4},-\frac{4\gamma\mu^2}{r_h^4}\right].\label{constU}
\end{equation}
The Hawking temperature for the charged Born-Infeld $AdS_4$ black hole is given by
\begin{equation}
T=\frac{1}{4\pi}\left[\frac{1}{r_h}+\left\lbrace\frac{\kappa^2}{3\gamma}+\frac{3}{L^2}\right\rbrace r_h-\frac{\kappa^2}{2r_h\gamma}\sqrt{r_h^4+4\gamma\mu^2}\right].
\end{equation} 
In order to obtain a  black hole solution with vector hair in the bulk it is necessary to consider solutions to the equations of motion (\ref{eom1}-\ref{eom5}) with a nontrivial profile for $\rho_{x}$ (corresponding to the $x$ component of $\rho_\mu$). However the full set of coupled equations of motion do not admit an analytic solution for this case . Hence these equations need to be solved numerically. From the equations of motion we observe that for a nontrivial solution we need to determine five independent functions $\Theta=\left\lbrace\phi,\rho_{x},f,h,\chi\right\rbrace$. For this suitable boundary conditions must be imposed at the conformal boundary $r\rightarrow\infty$ and at the horizon  $r=r_h$ of the $AdS_4$ bulk. The asymptotic form of the functions $\Theta=\left\lbrace\phi,\rho_{x},f,h,\chi\right\rbrace$ near the AdS boundary $r\rightarrow\infty$ may be written as,
\begin{eqnarray}
\phi &=& \mu-\frac{\rho}{r}+\cdots,~~~\rho_x=\frac{\rho_{x-}}{r^{\Delta_{-}}}+\frac{\rho_{x+}}{r^{\Delta_{+}}}+\cdots,\label{bdryfieldexp}\\
f &=& r^2(1+\frac{f_3}{r^3})+\cdots,~~~ h=1+\frac{h_3}{r^3}+\cdots,~~~ \chi =0+\frac{\chi_3}{r^3}\cdots,\label{bdrymetricexp}
\end{eqnarray}
where $\Delta_{\pm}=\frac{1\pm\sqrt{1+4 m^{2}}}{2}$ and the dots represent  higher order terms in powers of $1/r$. We require the vector condensate to arise spontaneously in the boundary theory, hence we impose the condition $\rho_{x-}=0$ in the asymptotic form of $\rho_x$ near the boundary. Here the coefficients $\mu,\rho$ and $\rho_{x+}$ represent the chemical potential, charge density and the $x$ component of the vacuum expectation value of the dual vector operator $<J_x>$ respectively in the dual boundary field theory.

For black hole configurations with regular event horizons we have $f(r_h)=0$ with the extra condition $\phi(r_h)=0$ so that  $g^{\mu\nu}A_{\mu}A_{\nu}$ term may remain finite at the horizon $r=r_h$. We also require the functions $\Theta=\left\lbrace\phi,\rho_{x},f,h,\chi\right\rbrace$ to be regular at the horizon $r=r_h$ which implies that all the functions must admit finite values and Taylor series expansions near the horizon as,
\begin{equation}
\Theta(r)=\Theta(r_h)+\Theta'(r_h)(r-r_h)+\cdots
\end{equation}
Analyzing the equations of motion and the horizon expansions of the functions, we may identify that there are five independent parameters  $\left\lbrace r_h,\phi'(r_h),\rho_{x}(r_h),h(r_h),\chi(r_h)\right\rbrace$ at the horizon.

It may be noted that our construction admits three useful scaling symmetries for the metric, bulk matter fields and the equations of the motion.  These symmetries may be listed as follows,
\begin{eqnarray}
e^{\chi}&\rightarrow & a^2 e^{\chi},~~~t\rightarrow a t,~~~\phi\rightarrow a^{-1}\phi, \label{sym1}\\
\rho_{x}&\rightarrow & a\rho_{x},~~~h\rightarrow a^2 h,~~~x\rightarrow a^{-1} x,~~~y\rightarrow a y,\label{sym2}\\
r &\rightarrow & a r,~~~f\rightarrow a^2 f,~~~(t,x,y)\rightarrow a^{-1}(t,x,y),~~~(\phi,\rho_{x})\rightarrow a (\phi,\rho_x),\label{sym3}
\end{eqnarray}
where a is a positive number. Now using the three symmetries described above,  near the horizon we set $\left\lbrace r_h=1,h(r_h)=1,\chi(r_h)=0\right\rbrace$ for our analysis. Similarly using the first two symmetries we set $\left\lbrace\chi(\infty)=0,h(\infty)=0\right\rbrace$ near the AdS boundary. Thus we are left with only two independent parameters at the horizon  $\left\lbrace \phi'(r_h),\rho_{x}(r_h)\right\rbrace$ and out of these two we will use one of them as a shooting parameter to obtain the source free condition $\rho_{x-}=0$. The remaining quantities like $\mu$,  $\rho$ etc. may be obtained by reading off the corresponding coefficients in the asymptotic forms as given in (\ref{bdryfieldexp}).

Finally under the third symmetry the quantities $\mu$, $\rho$, $T$, $\rho_{x+}$ and $\psi_{+}$ transform as,
\begin{equation}
\mu\rightarrow a \mu,~~~\rho\rightarrow a^2 \rho,~~~T\rightarrow a T,~~~\rho_{x+}\rightarrow a^{\Delta_{+} + 1}\rho_{x+}.
\end{equation}
Using the symmetries mentioned above we fix the chemical potential $\mu$ as the scaling parameter which implies that the finite temperature boundary field theory is considered to be in a grand canonical ensemble. One may then define the dimensionless scale invariant quantities in the grand canonical ensemble as,
\begin{equation}
 T/\mu,~~~(\rho_{x+})^\frac{1}{(\Delta_{+} + 1)}/\mu
\end{equation}
In next sections we will focus on the behavior of these scale invariant quantities which describe the relevant properties of the dual boundary field theory through the gauge/gravity duality.

\section{Gibbs free energy and dual stress energy tensor}

In order to establish which of the two phases (normal or condensed) is energetically favorable, we consider the behavior of the Gibbs free energy for the dual boundary field theory corresponding to the bulk configuration. Following the gauge/gravity dictionary one may note that the Gibbs free energy $(\Omega)$ of the boundary theory is related to the product of the on-shell Euclidean bulk action and the temperature $T$. In order to pose a well-defined, stationary Dirichlet variational problem one must add the Gibbons-Hawking boundary term \cite{PhysRevD.15.2752} to the Euclidean action and also a surface counter term for removing divergences.  The Euclidean action may be related to the Minkowski one by a minus sign as
\begin{equation}
-2\kappa^2 S_{Euclidean}=\int dx^4\sqrt{-g}\left({\cal R}+\frac{6}{L^2}+{\cal L}_{m}\right)+\int_{r\rightarrow\infty}dx^3\sqrt{-h}\left(2{\cal K}-\frac{4}{L^2}\right),\label{ActionE}
\end{equation}
where the $AdS$ length $L$ is set to unity for further analysis. Here, $h$ is the determinant of the induced metric on the AdS boundary, and ${\cal K}$ is the trace of the extrinsic curvature ${\cal K}_{\mu\nu}$. In order to relate the Gibbs free energy of the boundary theory $(\Omega)$ to the on-shell Euclidean bulk action, we begin by noting that there exists a relationship between the  matter Lagrangian and the stress energy tensor \cite{Hartnoll:2008kx} which may be described as,
\begin{equation}
T_{\mu\nu}=-g_{\mu\nu}{\cal L}_{m}.\label{rel1}
\end{equation}
The Einstein's equations (\ref{EinsEom}) may be written in the following fashion
\begin{eqnarray}
G_{\mu\nu}&=&\frac{1}{2}g_{\mu\nu}({\cal L}_{m}+6)=\frac{1}{2}g_{\mu\nu}({\cal L}_{tot}-{\cal R}),\label{rel2}
\end{eqnarray}
where, ${\cal L}_{tot}={\cal R}+6+{\cal L}_{m}$. Now considering the $xx$ and $yy$ component of the eq.(\ref{rel2}), we find that
\begin{equation}
{\cal L}_{tot}-{\cal R}=G_{~x}^{x}+G_{~y}^{y}.\label{rel3}
\end{equation}
One may also relate the Ricci scalar to the Einstein tensor as follows,
\begin{equation}
-{\cal R}=G_{~\mu}^{\mu}.\label{rel4}
\end{equation}
Using eq.(\ref{rel3}) and eq.(\ref{rel4}) with the ansatz eq.(\ref{metric}), we may rewrite the total Lagrangian in terms of the components of the Einstein tensor as follows,
\begin{equation}
{\cal L}_{tot}=-G_{~t}^{t}-G_{~r}^{r}=\frac{2f(r)-r f(r)\chi'(r)+2rf'(r)}{r^2}.\label{rel5}
\end{equation}
Thus plugging in the expressions for ${\cal L}_{tot}$ and ${\cal K}$ in eq.(\ref{ActionE}), we obtain the following expression for the on-shell Euclidean action,
\begin{equation}
-2\kappa^2 S_{Euclidean}^{on-shell}=\int dx^3 e^{-\frac{\chi}{2}}r\left(-2f-rf'+r\chi'f+4r\sqrt{f}\right)\vert_{r\rightarrow\infty}.\label{rel6}
\end{equation}
Furthermore following \cite{Hartnoll:2008kx}, one must add a counter term quadratic in the bulk vector field that provides a contribution of the form
\begin{equation}
S_{ct}=-\int dx^3(\alpha \rho_{x-} \rho_{x+})\vert_{r\rightarrow\infty}
\end{equation}
where $\alpha$ is a constant. We observe that it is possible to choose the asymptotic values $(\rho_{x-}=0,\rho_{x+}\neq0)$ in eq.(\ref{bdryfieldexp}), in order to implement the source free condition at the boundary or fix $(\rho_{x+}=0,\rho_{x-}\neq0)$ at the boundary. For our purpose at least one of the $\rho_{x\pm}=0$, hence this term does not contribute to the computation of the free energy.  On substituting the asymptotic expansions (\ref{bdrymetricexp}) into (\ref{rel6}), we obtain the expression for the Gibbs free energy $(\Omega)$ of the boundary theory as ,
\begin{equation}
\Omega= \frac{1}{\beta} S_{Euclidean}^{on-shell}=\frac{V_2}{2\kappa^2} f_3,\label{FreeEn}
\end{equation}
where, $\beta=1/T$ and $V_2=\int dx dy$. Now from eq.(\ref{nrmlphase}) we  note that for the normal phase with $h_3=\chi_3=0$, the expression for the free energy is \begin{eqnarray}
\Omega_{nrml}&=&-\frac{V_2}{2\kappa^2}M^2,\nonumber\\
M^2 &=&\frac{1}{6 r_h\gamma}\left(r_{h}^4 \left(6\gamma + \kappa^2\right)-\kappa^2 r_{h}^2 \sqrt{4\gamma\mu^2 + r_{h}^4}+ 8\gamma\kappa^2 \mu^2 {}_2 F_{1}\left[\frac{1}{4},\frac{1}{2},\frac{5}{4},-\frac{4\gamma\mu^2}{r_{h}^4}\right]\right)
\end{eqnarray}
here the mass of the black hole $(M)$ as a function of $r_h$ is obtained through solving the condition $f(r_h)=0$. As stated earlier, a non zero vacuum expectation value of the operator $J_x$ dual to the vector field $\rho_x$, breaks both the $U(1)$ gauge symmetry and the rotational invariance in $x-y$ plane. This may be established from the stress-energy tensor of the dual boundary field theory \cite{Balasubramanian:1999re} which is given as,
\begin{equation}
T_{a b}=\frac{1}{\kappa^2}\lim_{r\rightarrow\infty}[r\left({\cal K}h_{ab}-{\cal K}_{ab}-2h_{ab}\right)]
\end{equation}
where, $a,b=\{t,x,y\}$. Using the asymptotic expansions (\ref{bdrymetricexp}), we obtain the non-zero components of the stress-energy tensor for the dual boundary field theory as,
\begin{eqnarray}
T_{tt}&=&\frac{1}{\kappa^2}(-f_3),\nonumber\\
T_{xx}&=&\frac{1}{2\kappa^2}(-f_3+3h_3+3\chi_3),\nonumber\\
T_{yy}&=&\frac{1}{2\kappa^2}(-f_3-3h_3+3\chi_3),
\end{eqnarray}
For the normal phase we see that, $T_{xx}=T_{yy}$ and $\Omega/V_{2}=-T_{tt}$, as $h_3=0, \chi_3=0$. This indicates that the normal phase is isotropic in the $x-y$ plane. However for the superconducting phase the rotational invariance in the $x-y$ plane is broken due to the non-zero $<J_x>$, $h_3$ and $\chi_3$. This translates to the fact that, $T_{xx}\neq T_{yy}$ for the superconducting phase due to the anisotropic condensate. 
\section{Normal and the superconducting phase of the boundary field theory}
In this section we will focus on the superconducting phase of the boundary theory dual to our bulk model in the context of gauge/gravity duality. The normal phase of the boundary theory is dual to the charged Born-Infeld $AdS_4$ black hole in the bulk described by eq.(\ref{nrmlphase}). Whereas the superconducting phase of the boundary theory corresponds to the solution $\rho_x\neq 0$ of the equations (\ref{eom1}-\ref{eom5}). One can see that the bulk theory described here depends on three independent parameters namely the mass $(m)$ of the bulk vector field, the backreaction parameter $(\kappa)$ and the Born-Infeld parameter $(\gamma)$. We now proceed with computing the superconducting phase numerically for different range of values for these parameters. We will first focus on the behavior of the system with varying values of the mass (or the parameter $m^2$) of the bulk vector field. We observe that there exists a certain critical value of the parameter $m^2$ labeled as $m_c^2=0$, above which the condensed phase seems to exist for all temperatures below  $T_c$. Whereas for the case $m^2 < m_c^2$ the condensed phase cannot exist below a certain temperature $T_0$ which is less than the critical temperature $T_c$. This behavior was also observed for the case of p-wave superconductors described in \cite{Cai:2013aca} with $m_c^2$ also being zero. Now in order to compare our results with \cite{Cai:2013aca} we have considered the values for the mass as, $m^2=3/4$ for the case $m^2 > m_c^2$ and $m^2=-3/16$ for the case $m^2 < m_c^2$.
\subsection{$m^2=3/4$}
Now to begin with we assume $m^2=3/4$, as an example for the case $m^2> m^2_c$. For this case the normal phase of the boundary field theory exists above the critical temperature for all values of the backreaction parameter $\kappa$ and the Born-Infeld parameter $\gamma$. Whereas the p-wave superconducting phase of the boundary field theory exists only below a certain critical temperature $T_c$. As described earlier, the phase structure of the (2+1)-dimensional boundary field theory described here depends on the two parameters namely the backreaction parameter $\kappa$ and the Born-Infeld parameter $\gamma$ in the bulk. Through the numerical computations we find that the order of the phase transition may be changed from second order to first order by increasing the strength of the backreaction parameter for fixed but small values of the Born-Infeld parameter \footnote{It is to be noted that in the limit of small Born-Infeld parameter $\gamma$, the nonlinear Born-Infeld term introduced in the bulk reduces to a Maxwell term. Thus for very small values of  Born-Infeld parameter $\gamma$ the phase structure for our construction of p-wave holographic superconductors in the framework of a bulk Einstein-Born-Infeld theory matches with the phase behaviour observed for the case of p-wave holographic superconductors described in \cite{Cai:2013aca}}. Thus we obtain a critical value $(\kappa_c)$ of the backreaction parameter $\kappa$  such that the phase transition is of first order for $\kappa > \kappa_c$ and second order for $\kappa < \kappa_c$. Furthermore, we also observe that the phase transition may be changed from second order to first order by increasing the strength of the Born-Infeld parameter $\gamma$  for fixed values of the backreaction parameter $\kappa$ lying near the critical value $\kappa_c$. This indicates that  for $\kappa=\kappa_c$,  there exists a critical value ($\gamma_c$) of the Born-Infeld parameter such that the phase transition is of first order for $\gamma > \gamma_c$ and second order for $\gamma < \gamma_c$. For $m^2=3/4$, we obtain the critical values of the Born-Infeld parameter  and the backreaction parameter as $\gamma_c\simeq 0.03582$ and $\kappa_c \simeq 0.7364$ respectively.
\begin{figure}[H]
\centering
\begin{minipage}[b]{0.5\linewidth}
\includegraphics[width =2.5in,height=1.6in]{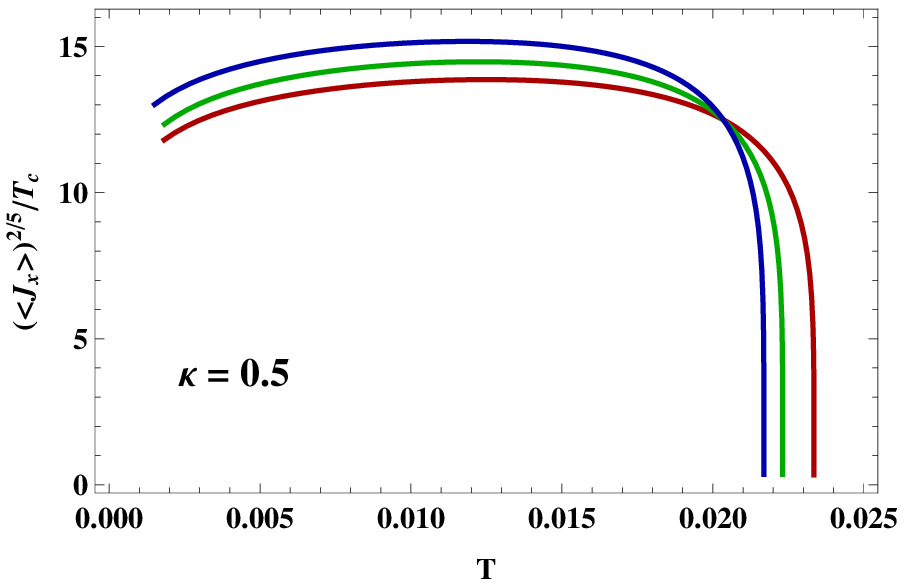}
\end{minipage}%
\begin{minipage}[b]{0.5\linewidth}
\includegraphics[width =2.5in,height=1.6in]{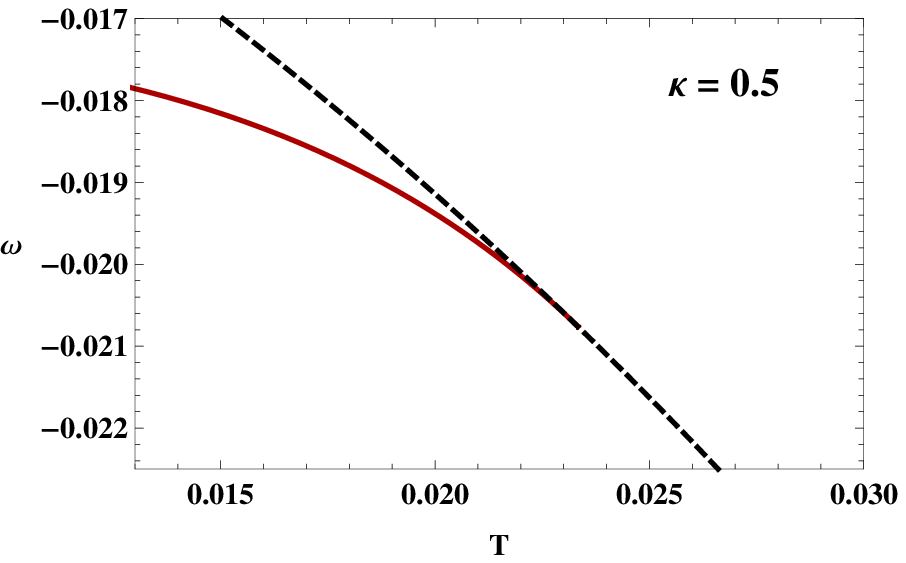}
\end{minipage}\quad
\caption{\label{fig:mass_3by4_condns1}Figures showing the condensate $(<J_x>)^{2/5}/T_c$ and the Free Energy $\omega=2\kappa^2\Omega/V_2$ in grand canonical ensemble as a function of temperature with the backreaction parameter fixed at $\kappa=0.5$. The red, green and the blue curves in left figure represent the different values of Born-Infeld parameter $\gamma=0.001,0.008$ and $0.01$ respectively and the red curve corresponds to the critical value of the temperature $T_c=0.004158\mu$.  For $\kappa=0.5$ and $\gamma=0.001$, The red curve in the right figure shows the free energy of the superconducting phase whereas the dashed black curve represents the normal phase. Here the mass of the vector field $\rho_x$ is taken to be $m^2=3/4$. } \end{figure} 

In fig.(\ref{fig:mass_3by4_condns1})  we have plotted the condensate $(<J_x>)^{2/5}/T_c$ as a function of temperature at fixed value of the backreaction parameter,  $\kappa=0.5<\kappa_c$  and  varying values of the Born-Infeld parameter $\gamma$. From the red curve corresponding to, $\kappa=0.5$ and $\gamma=0.001$  shown in the fig.(\ref{fig:mass_3by4_condns1}),  we observe that the dual vector operator $J_x$ acquires a non zero vacuum expectation value below a certain critical temperature $T_c\simeq0.0042$ and rises continuously from zero to saturation at temperatures well below $T_c$. Our numerical results also suggest that for $\kappa<\kappa_c$, the critical exponent for the phase transition  is always $1/2$  and  $<J_x>\sim (1-T/T_c)^{1/2}$,  which is a general mean field behaviour related to second order phase transitions. We have also plotted the grand potential  $(\omega=2\kappa^2\Omega/V_2)$ as a function of temperature  in the left plot of fig. (\ref{fig:mass_3by4_condns1}). From the free energy plot we observe that below the critical temperature $T_c$, the superconducting phase  has lower free energy than the normal phase.  This suggests that the superconducting phase is thermodynamically favored. From the condensate plots in fig.(\ref{fig:mass_3by4_condns1}),  we also observe that on increasing the value of the Born-Infeld parameter $\gamma$, the critical temperature drops to lower values  thereby making the condensate formation  harder.

\begin{figure}[H]
\centering
\begin{minipage}[b]{0.5\linewidth}
\includegraphics[width =2.5in,height=1.6in]{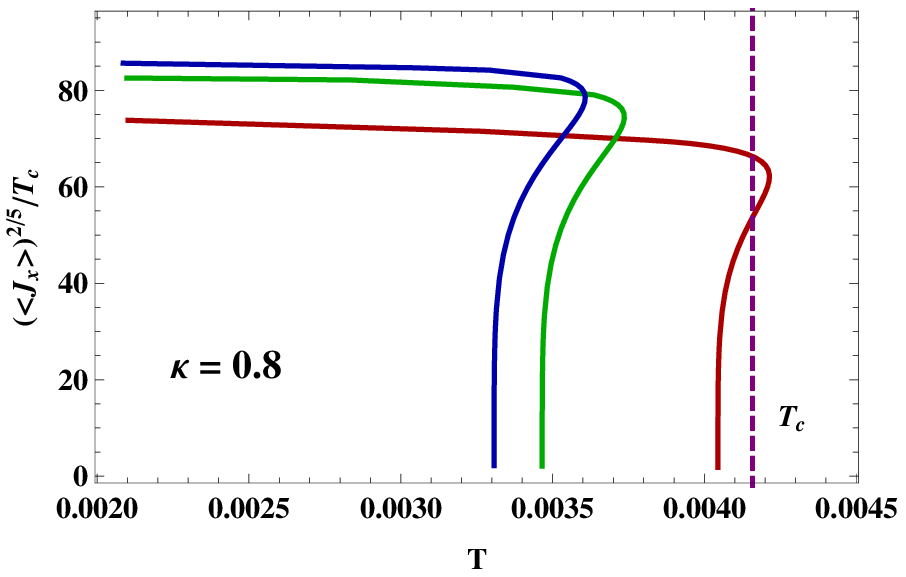}
\end{minipage}%
\begin{minipage}[b]{0.5\linewidth}
\includegraphics[width =2.5in,height=1.6in]{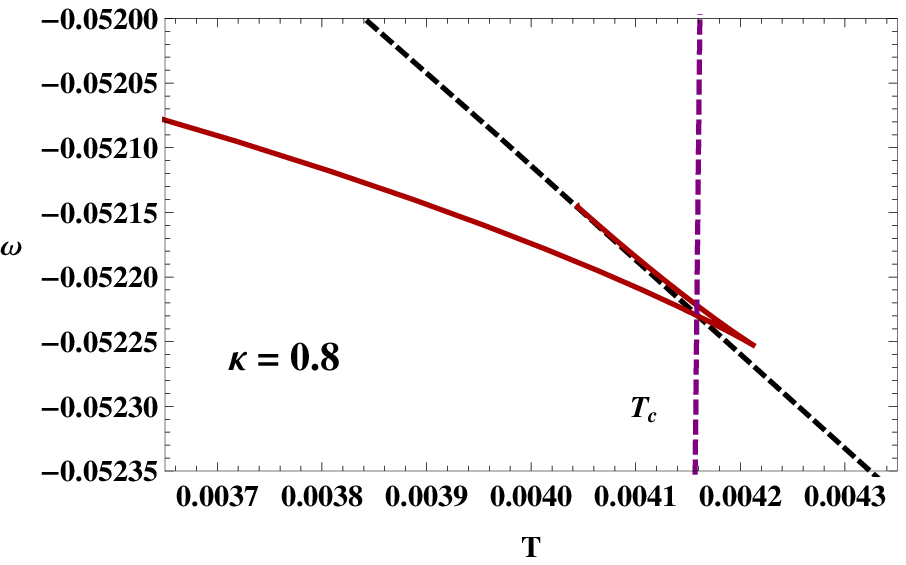}
\end{minipage}\quad
\caption{\label{fig:mass_3by4_condns2}Figures showing the condensate $(<J_x>)^{2/5}/T_c$ and the Free Energy $\omega=2\kappa^2\Omega/V_2$ in grand canonical ensemble as a function of temperature with the backreaction parameter fixed at $\kappa=0.8$. The red, green and the blue curves in left figure represent the different values of Born-Infeld parameter $\gamma=0.001,0.008$ and $0.01$ respectively and the red curve corresponds to the critical value of the temperature $T_c=0.004158\mu$. For $\kappa=0.8$ and $\gamma=0.001$, The red curve in the right figure shows the free energy of the superconducting phase whereas the dashed black curve represents the normal phase. Here the mass of the vector field $\rho_x$ is taken to be $m^2=3/4$. }
\end{figure}
\begin{figure}[H]
\centering
\begin{minipage}[b]{0.5\linewidth}
\includegraphics[width =2.5in,height=1.6in]{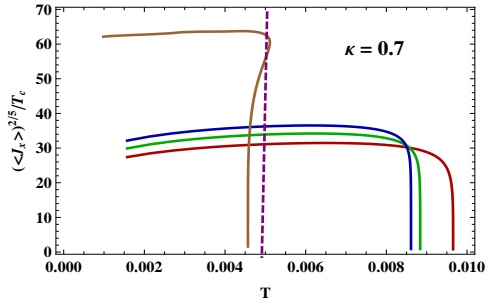}
\end{minipage}
\caption{\label{fig:mass_3by4_condns3} Figure showing the condensate $(<J_x>)^{2/5}/T_c$ as a function of temperature for varying values of the Born-Infeld parameter $\gamma$ with the backreaction parameter fixed at $\kappa=0.7$. The red, green and the blue curves represent the different values of Born-Infeld parameter $\gamma=0.001,0.008$ and $0.01$ respectively. The brown curve is for $\gamma=0.05$ with critical value of temperature, $T_c=0.004912\mu$ denoted by vertical dotted line. Here the mass of the vector field $\rho_x$ is taken to be $m^2=3/4$. } \end{figure} 

As we have observed, there exists a critical value $(\kappa_c)$ of the backreaction parameter, below which the formation of the condensate occurs through a second order phase transition.  We now focus on the other regime of the values for the backreaction parameter $\kappa$,  which lie above the critical value $\kappa_c$. In fig.(\ref{fig:mass_3by4_condns2}),  we have plotted the condensate $(<J_x>)^{2/5}/T_c$ as a function of temperature  at a fixed value of the backreaction parameter, 
$\kappa=0.8>\kappa_c$  and different values of the Born-Infeld parameter $\gamma$. It may be observed from the figure that the condensate is now multiple valued  at the critical temperature $T_c\simeq 0.0042$. We observe that there are two new sets of branches for the condensate at temperatures ( e.g.  $T \simeq 1.02T_c$)  below than the critical temperature. The upper-branch corresponds to  large values of $<J_x>$ whereas the lower branch  is for  small  values of  $<J_x>$.  In order to determine which of these branches correspond to the thermodynamically favored phase,  we have plotted the grand potential $(\omega=2\kappa^2\Omega/V_2)$ as a function of temperature in the left plot of the fig.(\ref{fig:mass_3by4_condns2}). It is observed from the plot that above the $T_c\simeq 0.0042$, the free energy develops a  {\it swallow tail},  which is a typical characteristic for  first order phase transitions.  In contrast the free energy for the normal phase $(<J_x>=0)$  shows that it is the preferred phase at high temperatures. As the temperature is lowered to $T_c$, the upper-branch corresponding to the condensed phase with large values of $<J_x>$ now represents the thermodynamically favored phase. Similar to the case of $\kappa<\kappa_{c}$ the critical temperature drops to lower values as the value of the Born-Infeld parameter $\gamma$ is increased
thereby making the condensation harder.

\begin{figure}[H]
\centering
\begin{minipage}[b]{0.5\linewidth}
\includegraphics[width =2.5in,height=1.6in]{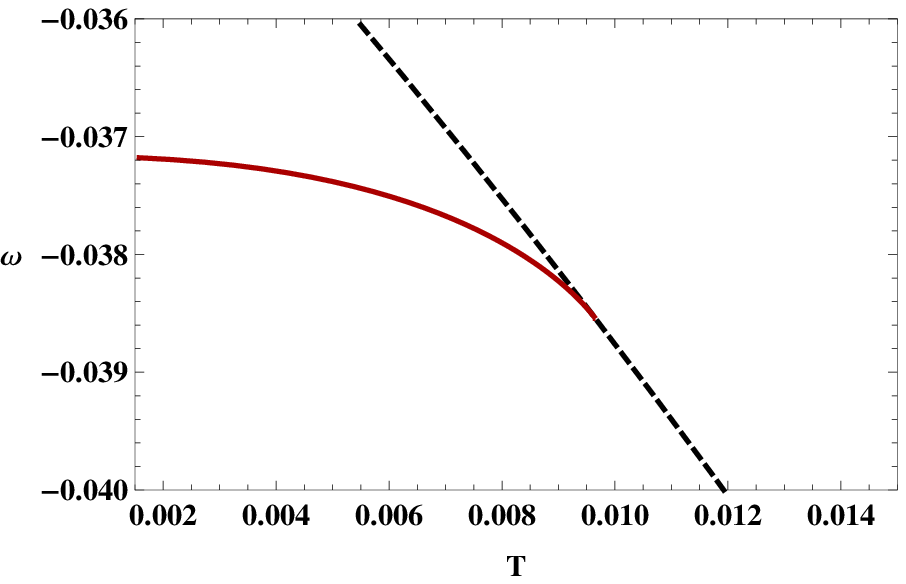}
\end{minipage}%
\begin{minipage}[b]{0.5\linewidth}
\includegraphics[width =2.5in,height=1.6in]{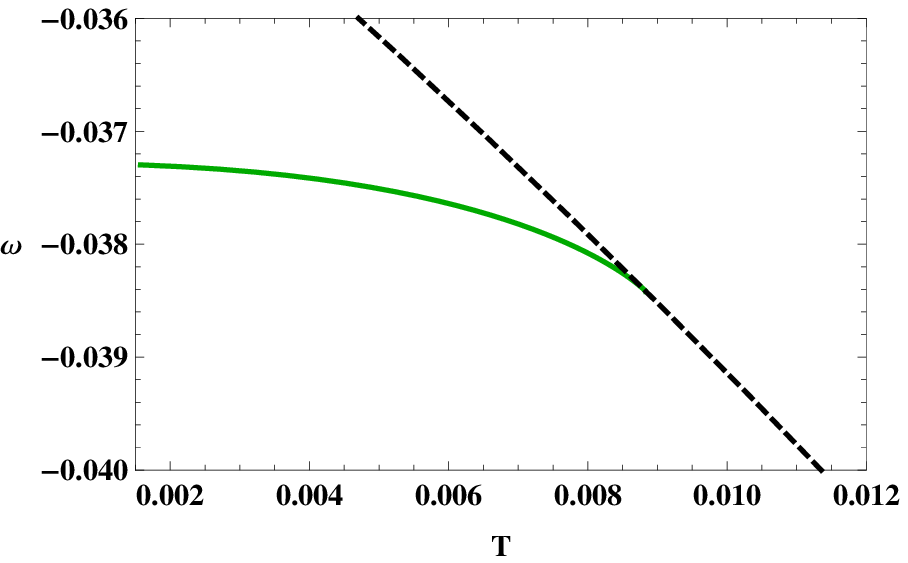}
\end{minipage}\quad
\begin{minipage}[b]{0.5\linewidth}
\includegraphics[width =2.5in,height=1.6in]{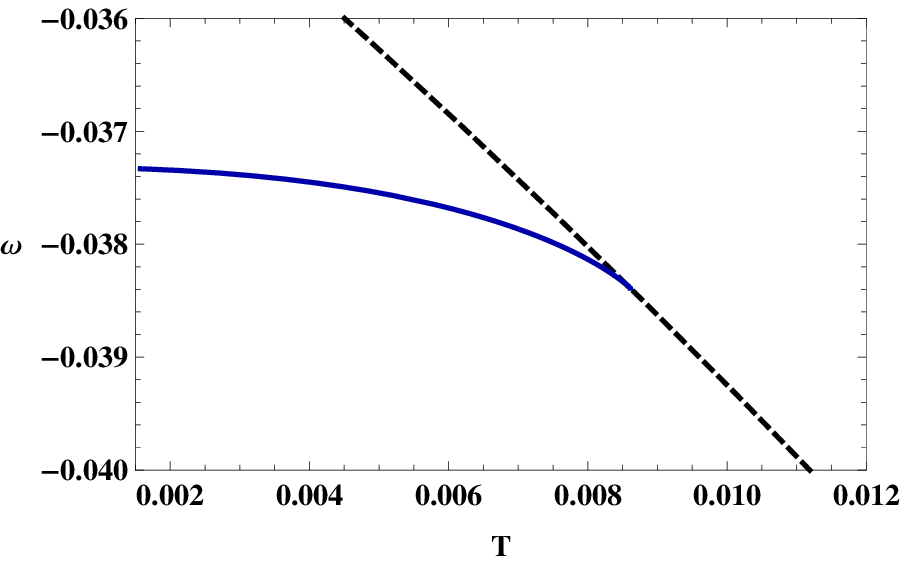}
\end{minipage}%
\begin{minipage}[b]{0.5\linewidth}
\includegraphics[width =2.5in,height=1.6in]{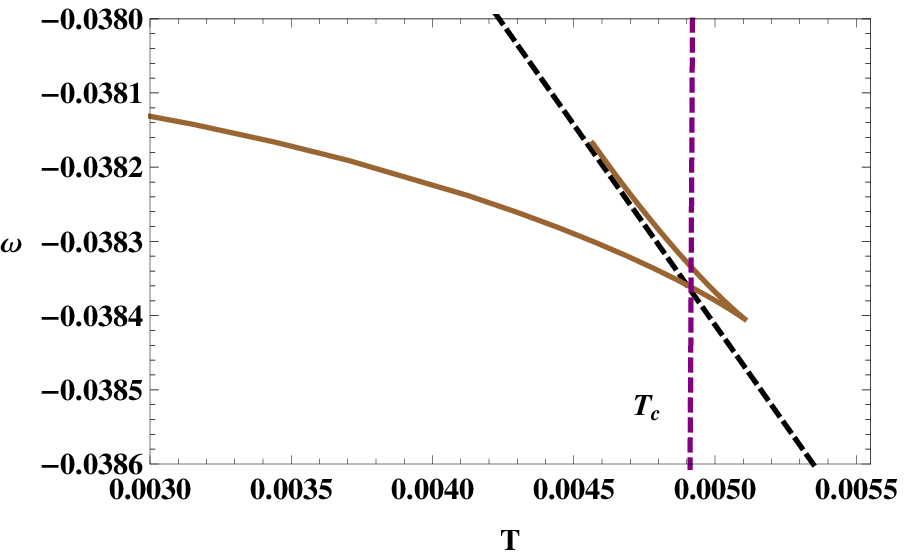}
\end{minipage}%
\caption{\label{fig:mass_3by4_FreeEn2} Plots showing the Free Energy $\omega=2\kappa^2\Omega/V_2$ in grand canonical ensemble with respect to the temperature. Here the plots are drawn for fixed value of the backreaction parameter $\kappa=0.7$ and for varying values of the Born-Infeld parameter $\gamma=0.001,0.008,0.01$ and $0.05$ represented by the red, green, blue and the brown curves respectively.}
\end{figure}
The interesting case arises when  we consider the value of the backreaction parameter near the critical value $(\kappa=0.7\simeq\kappa_c)$ . In the fig.(\ref{fig:mass_3by4_condns3}) and fig.(\ref{fig:mass_3by4_FreeEn2}),  we have plotted the condensate $(<J_x>)^{2/5}/T_c$ and the grand potential  $(\omega=2\kappa^2\Omega/V_2)$  as a function of temperature at the critical value of the backreaction parameter for different values of the Born-Infeld parameter $\gamma$. From both the figures it is observed that as the value of the Born-Infeld parameter is increased, the condensate changes its behavior from being single valued to being multiple valued implying a change in the order of the superconducting phase transition. 
This shows that  there exists a critical value $(\gamma_c\simeq 0.036)$ for the Born-Infeld parameter at which the phase transition changes from second order to first order.

\begin{figure}[H]
\centering
\begin{minipage}[b]{0.5\linewidth}
\includegraphics[width =2.5in,height=1.6in]{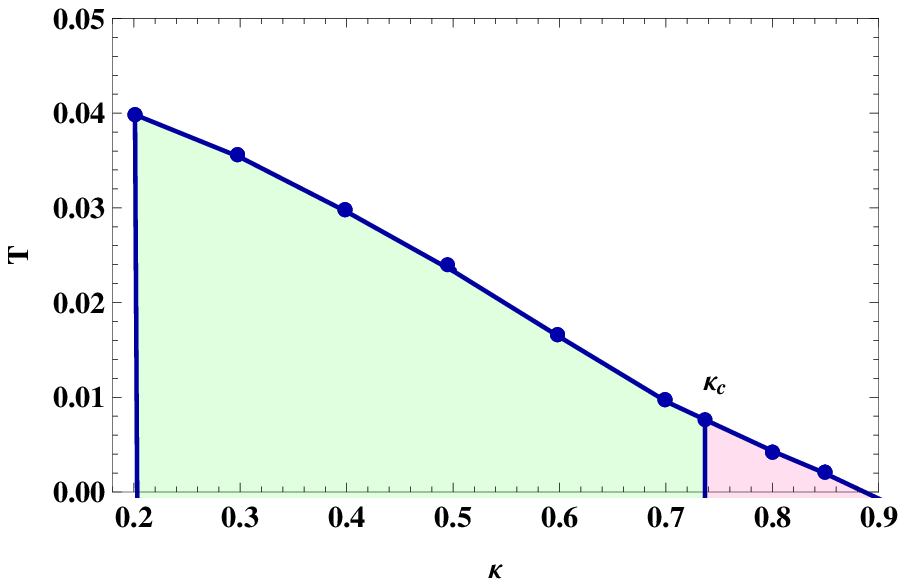}
\end{minipage}%
\begin{minipage}[b]{0.5\linewidth}
\includegraphics[width =2.5in,height=1.6in]{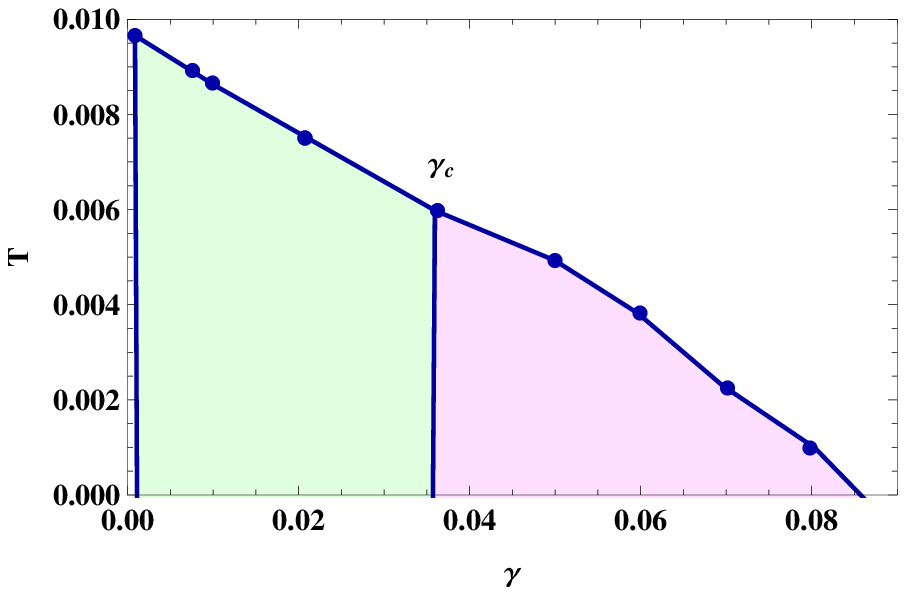}
\end{minipage}\quad
\caption{\label{fig:Phase_mass_3by4} The phase diagram for $m^2 = 3/4$. The left graph shows the $(T,\kappa)$ phase diagram at fixed $\gamma=0.001$ where the solid curve separates the condensed phase from the normal phase. The critical value $\kappa_c$ divides the condensed phase into two parts. The case $\kappa<\kappa_c$ is associated with second order phase transition (green area), while $\kappa > \kappa_c$ corresponds to first order transition (red area). Whereas, The right graph shows the $(T,\gamma)$ phase diagram at fixed $\kappa=0.07\simeq\kappa_c$ with the solid curve separating the condensed phase from the normal phase. The critical value $\gamma_c=0.0358$ divides the condensed phase into two parts. The case $\gamma<\gamma_c$ is associated with second order phase transition (green area), while $\gamma > \gamma_c$ corresponds to first order transition (red area).}
\end{figure}

The main results of this subsection are summarized in the $(T,\kappa)$ and $(T,\gamma)$ phase diagrams shown in the fig.(\ref{fig:Phase_mass_3by4}). The solid curves in both the phase diagrams represent the critical temperature $T_c$ for the superconducting phase transition. Depending on the values of $\kappa$ or $\gamma$, the region below the curve represents the superconducting phase of the boundary theory whereas the region above the curve corresponds to the normal or the metastable phase of the boundary field theory. The $(T,\kappa)$ phase diagram corresponding to fixed $\gamma=0.001$, shows the existence of a critical value  $(\kappa_c)$  of the backreaction parameter $\kappa$, above which the phase transition is of second order, whereas below it the phase transition is of first order. Similarly  the $(T,\gamma)$  phase diagram at fixed $\kappa=0.7\simeq\kappa_c$ shows the existence of a critical value $(\gamma_c)$  of the Born-Infeld parameter $\gamma$, below which the condensate formation occurs through a second order phase transition whereas above it the condensate formation is through a first order phase transition. The phase diagrams also describe that for increasing values of $\kappa$ or $\gamma$ the critical temperature $T_c$ decreases gradually implying that the superconducting phase transition is hindered for large values of both the backreaction parameter $\kappa$ and the Born-Infeld parameter $\gamma$ .

\subsection{$m^2=-3/16$}
In this section we now consider the case for $m^2<m^2_c$ to study the behavior of the condensate for various possible values of the backreaction parameter $\kappa$ and the Born-Infeld parameter $\gamma$. As a particular example we consider the mass of the vector field to be $m^2=-3/16$. Through the numerical computations we observe that for a small but fixed value of the Born-Infeld parameter $\gamma$, there exist two curious values of the backreaction parameter namely $\kappa_1$ and $\kappa_2$, such that the parameter space of $\kappa$ is partitioned into three regions $\kappa <\kappa_1$, $\kappa_1<\kappa<\kappa_2$ and $\kappa_2<\kappa$ respectively. For a fixed value of the backreaction parameter $\kappa$ in  the three distinct regions mentioned above, the condensate  behaves differently for varying values of the Born-Infeld parameter $(\gamma)$.  This indicates the existence of certain bounds on the possible values of $\gamma$. We will elaborate more on these bounds for the Born-Infeld parameter $(\gamma)$ later. Thus for $m^2<m^2_c$, we  observe that depending on the values of the parameters $\kappa$ and $\gamma$, the thermodynamic behavior changes dramatically and the condensate formation may occur via a first order, second order, zeroth order \cite{Maslov:2004ap} or a retrograde phase transition \cite{Narayanan1994135}. 
\begin{figure}[H]
\centering
\begin{minipage}[b]{0.5\linewidth}
\includegraphics[width =2.5in,height=1.6in]{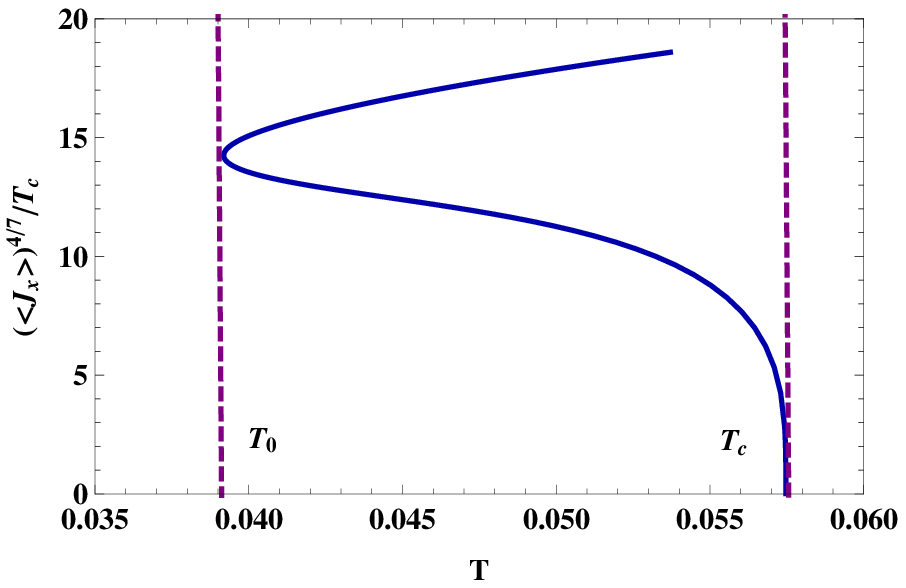}
\end{minipage}%
\begin{minipage}[b]{0.5\linewidth}
\includegraphics[width =2.5in,height=1.6in]{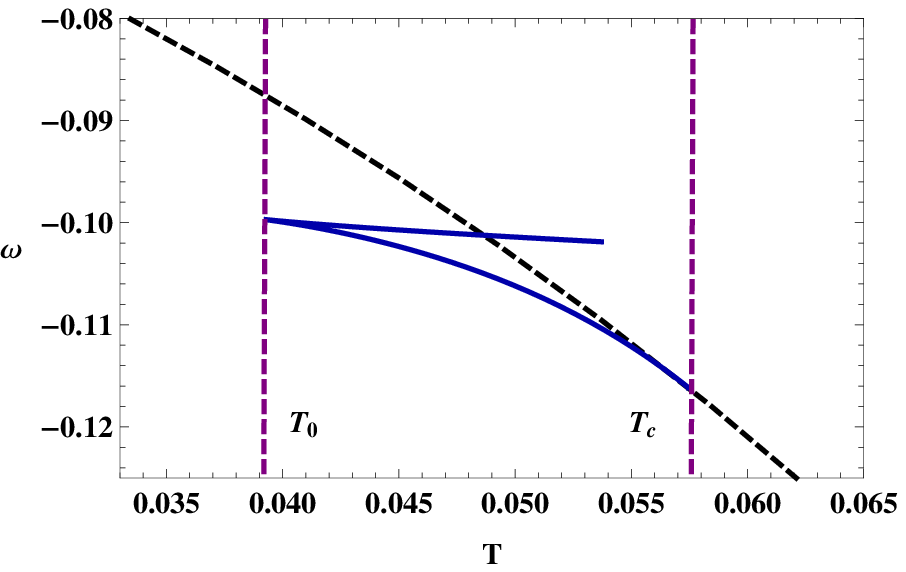}
\end{minipage}\quad
\caption{\label{fig:mass_3by16_condns1}Figures showing the condensate $(<J_x>)^{4/7}/T_c$ and the Free Energy $\omega=2\kappa^2\Omega/V_2$ in grand canonical ensemble as a function of temperature with  $\kappa=0.8$ and $\gamma=0.001$. Here the critical temperature $T_c=0.0575\mu$ and the temperature $T_0=0.0391$ are represented by the vertical dashed lines. For $\kappa=0.8$ and $\gamma=0.001$, the blue curve in the right figure shows the free energy of the superconducting phase whereas the dashed black curve represents the normal phase. Here the mass of the vector field $\rho_x$ is taken to be $m^2=-3/16$.} \end{figure} 

For the mass, $m^2=-3/16$ of the vector field considered above, we have obtained the critical values of $\kappa_1$ and $\kappa_2$ as $0.9828$ and $1.0485$ respectively for the Born-Infeld parameter $\gamma=0.001$. Now in order to study the behavior of the condensate in the region $\kappa<\kappa_1$, we have considered the value $\kappa=0.8$ of the backreaction parameter for which we have plotted the condensate $(<J_x>)^{4/7}/T_c$ as a function of temperature in the fig.(\ref{fig:mass_3by16_condns1}). From the graph it is observed that the condensate formation occurs via a second order phase transition at the critical temperature $T_c=0.0575\mu$ and the condensate formation ceases below a certain temperature, $T_0=0.0391\mu< T_c$. It is also to be noted that the condensate becomes multiple valued at the temperature $T_0$, with the upper branch representing larger values of $<J_x>$ and the lower branch representing smaller values. Furthermore for this case, in the left graph of the fig.(\ref{fig:mass_3by16_condns1}) we have plotted the grand potential $(\omega=2\kappa^2\Omega/V_2)$ as a function of temperature which shows a ``swallow tail" like behaviour near $T_0$ and a kink near $T_c$ indicating a second order phase transition. From the free energy plot it is observed that only the lower branch of the condensate with smaller values of $<J_x>$ , represents the thermodynamically favored phase in the temperature range $T_0<T<T_c$. Whereas the normal phase with $<J_x>=0$, is the preferred phase for the regions $T<T_0$ and $T>T_c$. The interesting feature that arises in the free energy plot is that the free energy now shows a discontinuity at $T_0$, which indicates a zeroth order phase transition. It is to be noted that in the context of superfluidity and superconductivity, a discontinuity of the free energy is related to a zeroth phase transitions that was first discovered  in an exactly solvable model described in \cite{Maslov:2004ap}.

\begin{figure}[H]
\centering
\begin{minipage}[b]{0.5\linewidth}
\includegraphics[width =2.5in,height=1.6in]{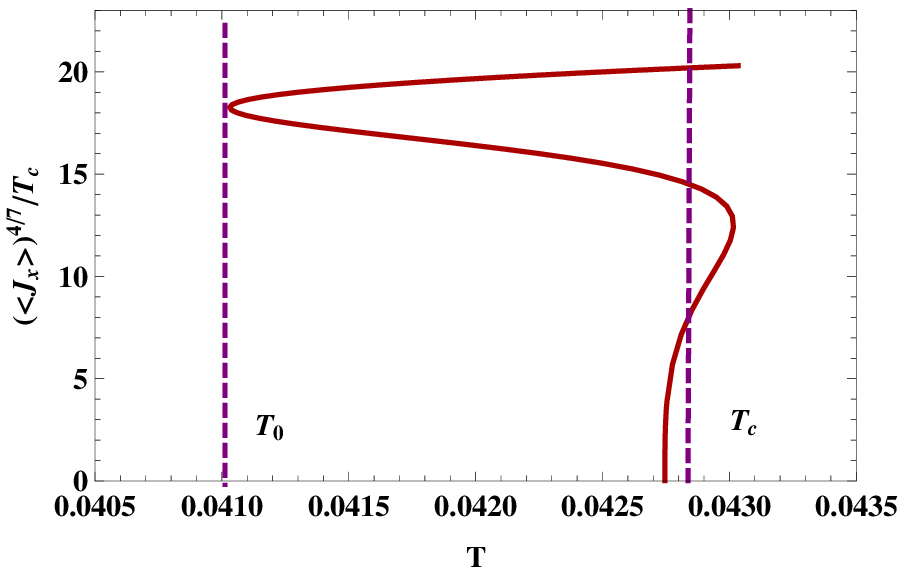}
\end{minipage}%
\begin{minipage}[b]{0.5\linewidth}
\includegraphics[width =2.5in,height=1.6in]{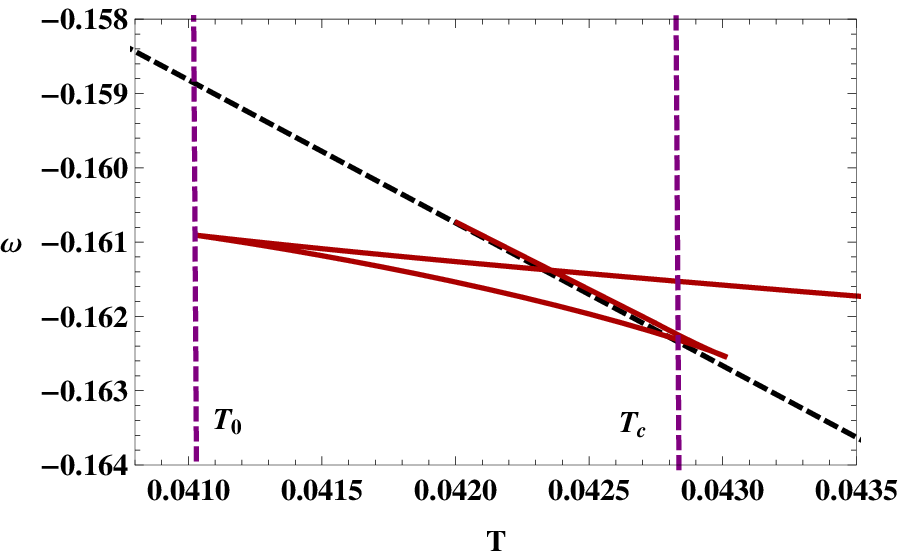}
\end{minipage}\quad
\caption{\label{fig:mass_3by16_condns2} Figures showing the condensate $(<J_x>)^{4/7}/T_c$ and the Free Energy $\omega=2\kappa^2\Omega/V_2$ in grand canonical ensemble as a function of temperature with  $\kappa=1.01$ and $\gamma=0.001$. The critical temperature is given by $T_c=0.0428\mu$ and the temperature $T_0=0.041$ are represented by the vertical dashed lines.  For $\kappa=1.01$ and $\gamma=0.001$, the the red curve in the right figure shows the free energy of the superconducting phase whereas the dashed black curve represents the normal phase. Here the mass of the vector field $\rho_x$ is taken to be $m^2=-3/16$. } \end{figure} 
\begin{figure}[H]
\centering
\begin{minipage}[b]{0.5\linewidth}
\includegraphics[width =2.5in,height=1.6in]{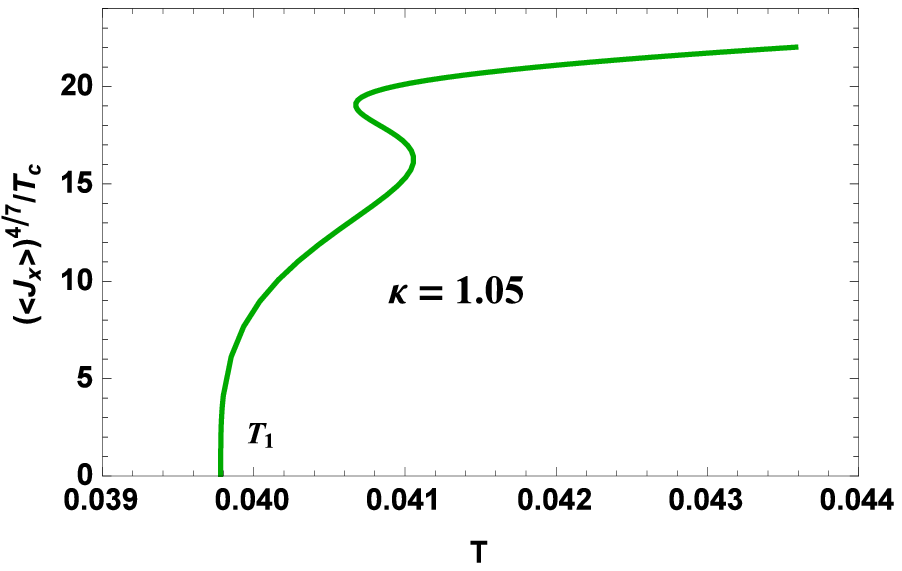}
\end{minipage}%
\begin{minipage}[b]{0.5\linewidth}
\includegraphics[width =2.5in,height=1.6in]{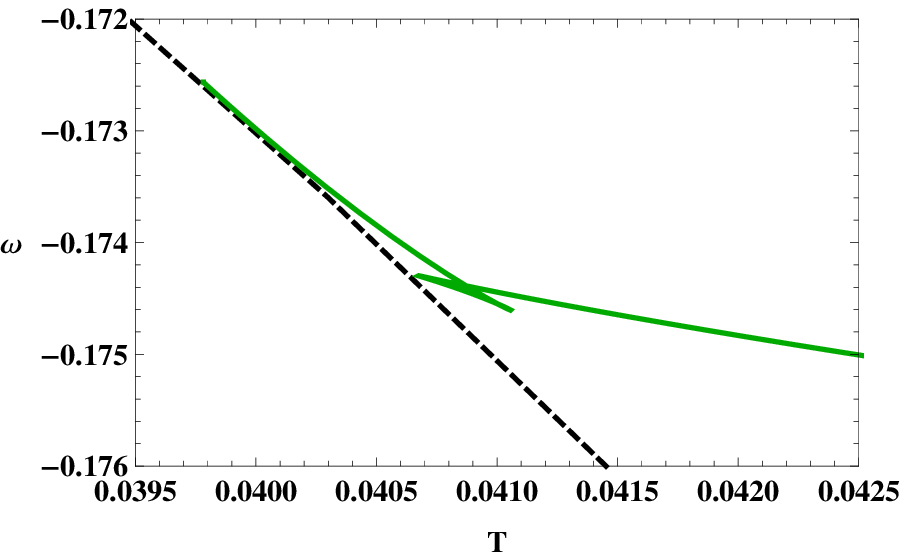}
\end{minipage}\quad
\caption{\label{fig:mass_3by16_condns3}Figures showing the condensate $(<J_x>)^{4/7}/T_c$ and the Free Energy $\omega=2\kappa^2\Omega/V_2$ in grand canonical ensemble as a function of temperature with  $\kappa=1.05$ and $\gamma=0.001$. The critical temperature above which the condensate appears is given by $T_1=0.0398\mu$. For $\kappa=1.05$ and $\gamma=0.001$, the the green curve in the right figure shows the free energy of the superconducting phase whereas the dashed black curve represents the normal phase. Here the mass of the vector field $\rho_x$ is taken to be $m^2=-3/16$. } \end{figure}
\begin{figure}[H]
\centering
\begin{minipage}[b]{0.5\linewidth}
\includegraphics[width =2.5in,height=1.6in]{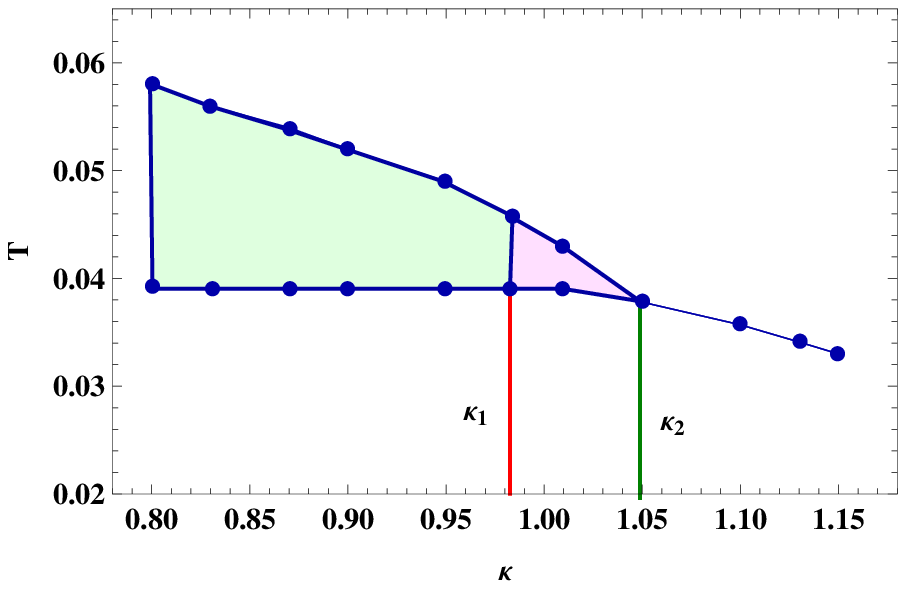}
\end{minipage}%
\begin{minipage}[b]{0.5\linewidth}
\includegraphics[width =2.5in,height=1.6in]{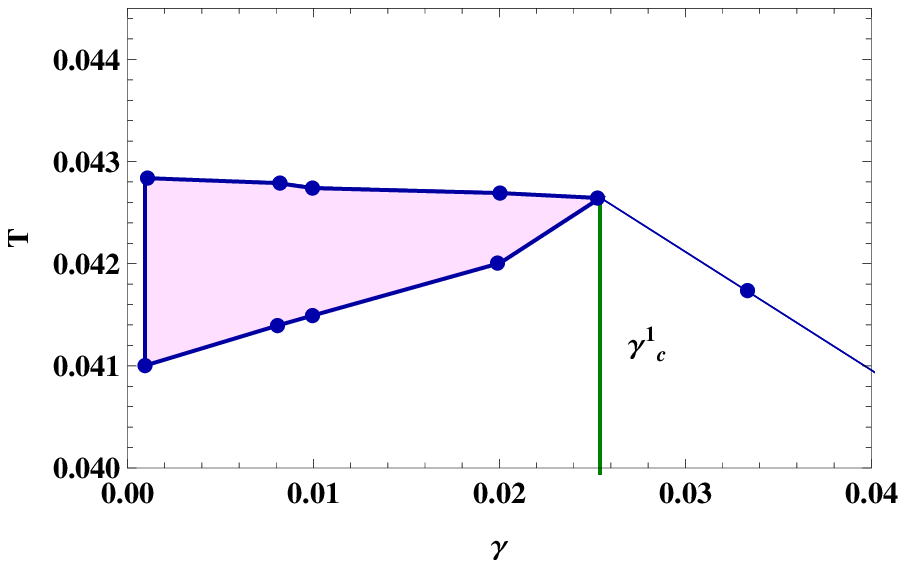}
\end{minipage}\quad
\caption{\label{fig:Phase_mass_3by16} The phase diagram for $m^2 = -3/16$. The left graph shows the $(T,\kappa)$ phase diagram at fixed $\gamma=0.001$ where the colored region between the $T_0$ curve (bottom) and the $T_c$ curve (Top) corresponds to the condensed phase. The critical values $\kappa_1$ and $\kappa_2$ divides the condensed phase into three parts. The case $\kappa<\kappa_1$ is associated with second order phase transition (green area), while $\kappa_1 <\kappa <\kappa_2$ corresponds to first order transition (red area). Whereas, The $T_0$ curve (bottom) and the $T_c$ curve (Top) merge with each other at $\kappa_2=1.0485$ resulting in the retrograde transition curve for $\kappa>\kappa_2$. For fixed $\kappa=1.01$, the critical value $\gamma^1_c=0.0254$ divides the condensed phase into two parts. The case $\gamma<\gamma^1_c$ is associated with first order phase transition (red area bounded between the $T_0$ curve (bottom) and the $T_c$ curve (Top)), while curve corresponding to $\gamma>\gamma^1_c$ corresponds to retrograde phase transition .}
\end{figure}

 
We now focus on the case with the backreaction parameter lying in the range,  $\kappa_1<\kappa<\kappa_2$ for $\gamma=0.001$. For this we have fixed the value of the backreaction parameter at, $\kappa=1.01$. In the left graph of the fig.(\ref{fig:mass_3by16_condns2}) we have plotted the condensate $(<J_x>)^{4/7}/T_c$ as a function of temperature for the value $\kappa=1.01$, of the backreaction parameter with the value $\gamma=0.001$, of the Born-Infeld parameter. From the plot it may  be observed that the condensate now becomes multiple valued at both the temperatures $T_0=0.041\mu$ and $T_c=0.0428\mu$ indicating a zeroth order phase transition at $T_0$ and a first order phase transition at $T_c$. Above $T_0$ the condensate bifurcates into an upper branch corresponding to larger values of $<J_x>$ and a lower branch corresponding to smaller values. Whereas near the critical temperature $T_c$, the condensate has three branches namely an upper branch with higher values of $<J_x>$, a lower branch with lower values of $<J_x>$ and a middle branch with values of $<J_x>$ lying between the other two. From the free energy verses temperature plot in the right side of the fig.(\ref{fig:mass_3by16_condns2}),  it is observed that at higher temperatures the normal phase with $<J_x>=0$ is the thermodynamically favored phase. However as the temperature is lowered to $T_c=0.0428\mu$, the free energy shows a ``swallow tail" behaviour indicating a first order phase transition and the thermodynamically favored phase is now represented by the middle branch. As we further lower the temperature the free energy shows a discontnuity at the temperature $T_0$ indicating a zeroth order phase transition. Through the numerics,  we observe that as the backreaction parameter is increased then both the critical temperature $(T_c)$ and the temperature $T_0$ start decreasing and finally $T_c$ coincides with $T_0$ at $\kappa=\kappa_2$ (see fig.(\ref{fig:Phase_mass_3by16})).
\begin{figure}[H]
\centering
\begin{minipage}[b]{0.5\linewidth}
\includegraphics[width =2.5in,height=1.6in]{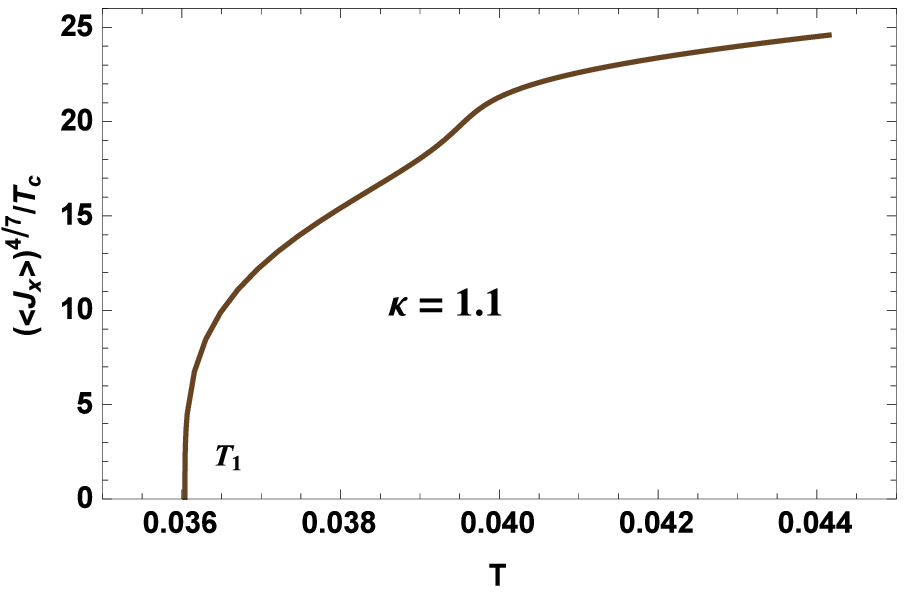}
\end{minipage}%
\begin{minipage}[b]{0.5\linewidth}
\includegraphics[width =2.5in,height=1.6in]{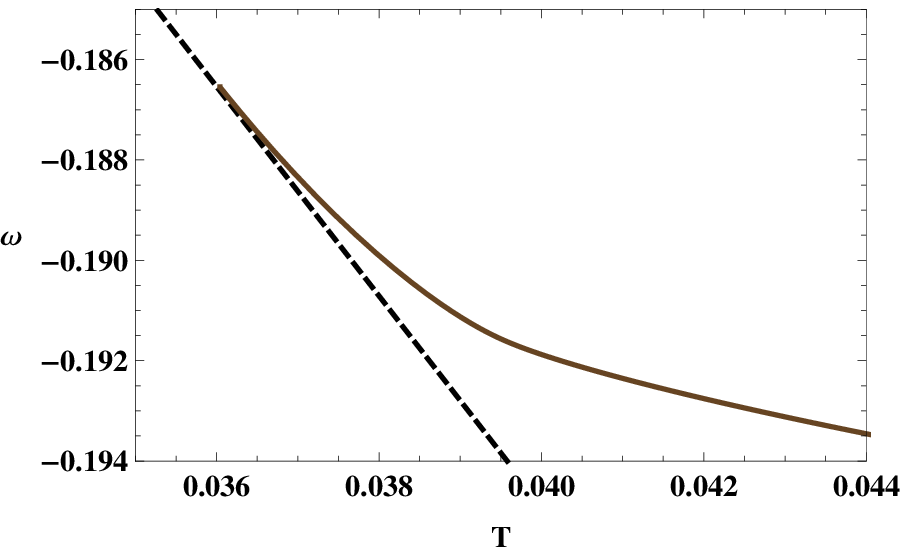}
\end{minipage}\quad
\caption{\label{fig:mass_3by16_condns4}Figures showing the condensate $(<J_x>)^{4/7}/T_c$ and the Free Energy $\omega=2\kappa^2\Omega/V_2$ in grand canonical ensemble as a function of temperature with  $\kappa=1.1$ and $\gamma=0.001$. The critical temperature above which the condensate appears is given by $T_1=0.036\mu$.  For $\kappa=1.1$ and $\gamma=0.001$, the the brown curve in the right figure shows the free energy of the superconducting phase whereas the dashed black curve represents the normal phase. Here the mass of the vector field $\rho_x$ is taken to be $m^2=-3/16$.  } \end{figure} 
\begin{figure}[H]
\centering
\begin{minipage}[b]{0.5\linewidth}
\includegraphics[width =2.5in,height=1.6in]{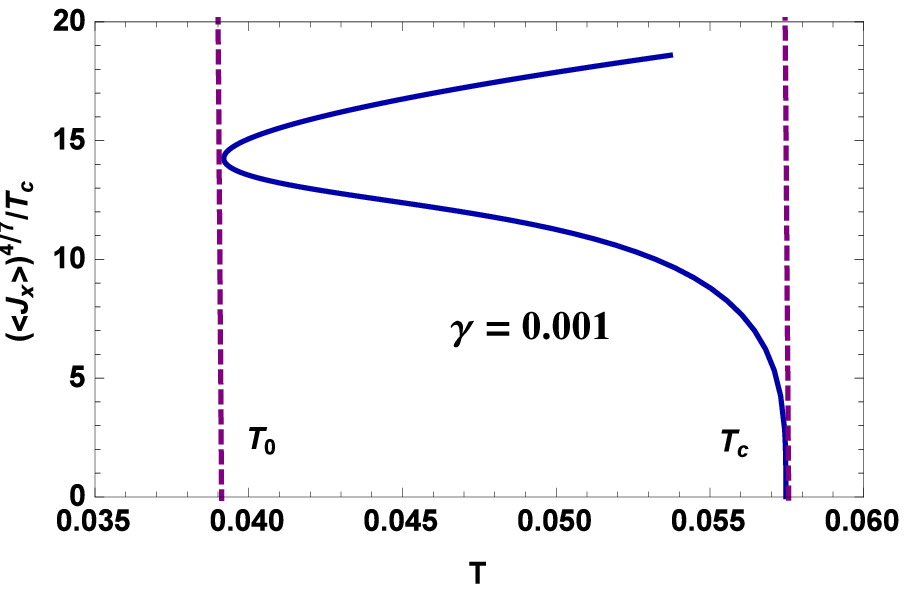}
\end{minipage}%
\begin{minipage}[b]{0.5\linewidth}
\includegraphics[width =2.5in,height=1.6in]{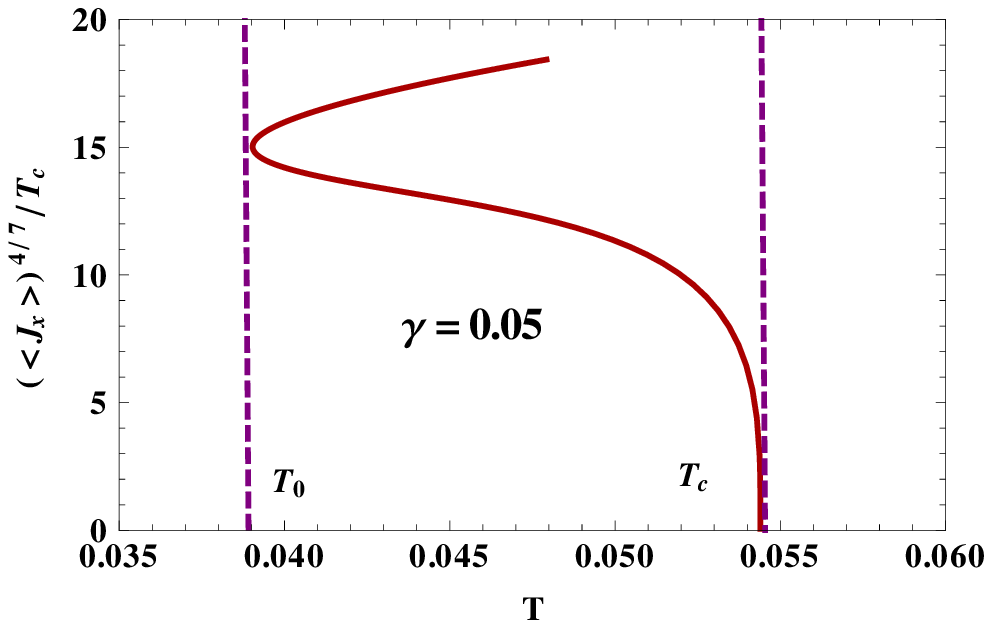}
\end{minipage}\quad
\begin{minipage}[b]{0.5\linewidth}
\includegraphics[width =2.5in,height=1.6in]{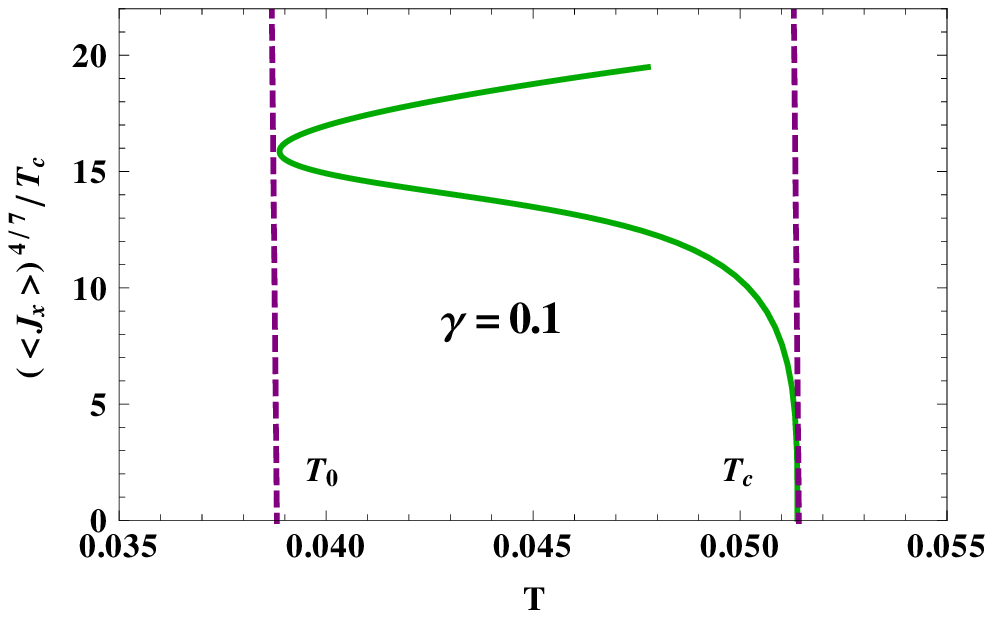}
\end{minipage}%
\caption{\label{fig:mass_3by16_condns5} Figures showing the condensate $(<J_x>)^{4/7}/T_c$ as a function of temperature for varying values of the Born-Infeld parameter $\gamma$ with the backreaction parameter fixed at $\kappa=0.8$. The blue, red and the green curves represent the different values of Born-Infeld parameter $\gamma=0.001,0.05$ and $0.1$ respectively. Here the mass of the vector field $\rho_x$ taken to be $m^2=-3/16$. } \end{figure} 
 Further increasing the backreaction parameter beyond the value $\kappa_2$, the condensate undergoes a drastic change in its behavior. For the case $\kappa > \kappa_2$, we have plotted the condensate $(<J_x>)^{4/7}/T_c$ as a function of temperature in fig.(\ref{fig:mass_3by16_condns3}) and fig.(\ref{fig:mass_3by16_condns4}) for the values of the backreaction parameter, $\kappa=1.05$ and $\kappa=1.1$ respectively with the value of the Born-Infeld parameter fixed at, $\gamma=0.001$. From the figures it  may be observed that the condensate now exists only above a certain temperature $T_1$ and it asymptotically increases with increasing values of  the temperature. The condensate remains multiple valued for the values of $\kappa$ lying near  $\kappa_2$ but as the value of $\kappa$ is increased further, the condensate becomes single valued again. Here the phenomena of the condensate existing above a certain temperature which is opposite to the general behavior of the condensate arising below a certain critical temperature, corresponds to the  black hole with ``vector hair" that has higher free energy than the normal phase as shown in right side of the fig.(\ref{fig:mass_3by16_condns3}) and fig.(\ref{fig:mass_3by16_condns4}). Thus these hairy black hole configurations correspond to metastable states in the boundary field theory. This phenomenon of a thermodynamically unfavorable phase existing above a certain critical temperature is known as {\it retrograde condensation} \cite{Narayanan1994135}.
\begin{figure}[H]
\centering
\begin{minipage}[b]{0.5\linewidth}
\includegraphics[width =2.5in,height=1.6in]{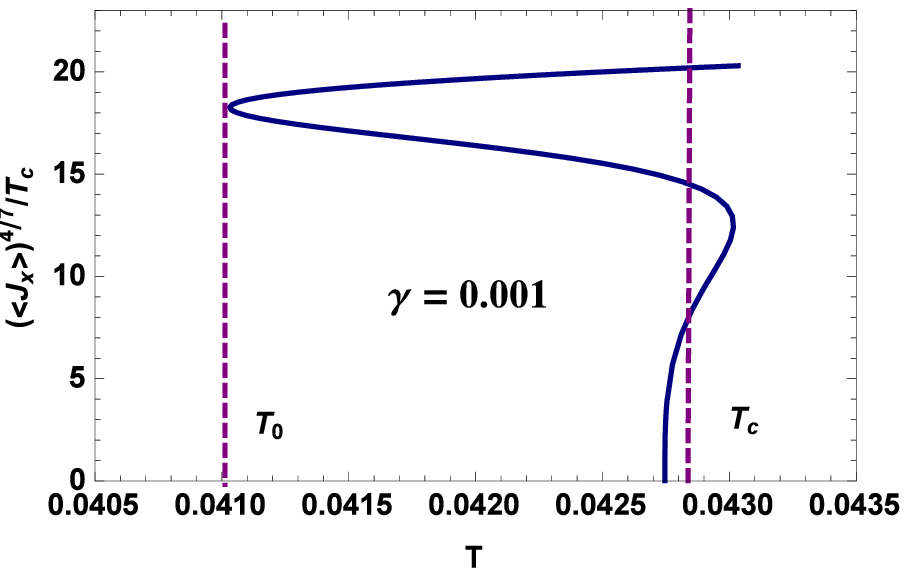}
\end{minipage}%
\begin{minipage}[b]{0.5\linewidth}
\includegraphics[width =2.5in,height=1.6in]{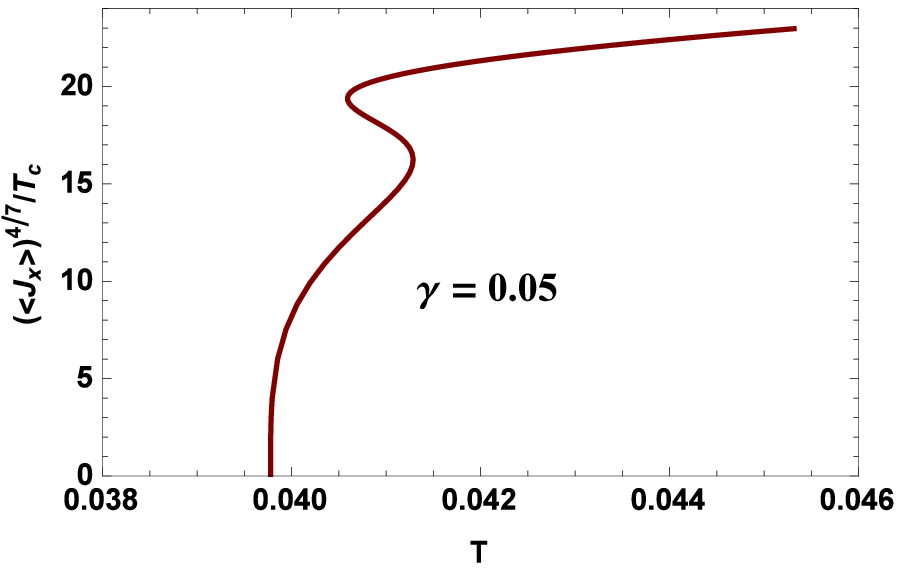}
\end{minipage}\quad
\begin{minipage}[b]{0.5\linewidth}
\includegraphics[width =2.5in,height=1.6in]{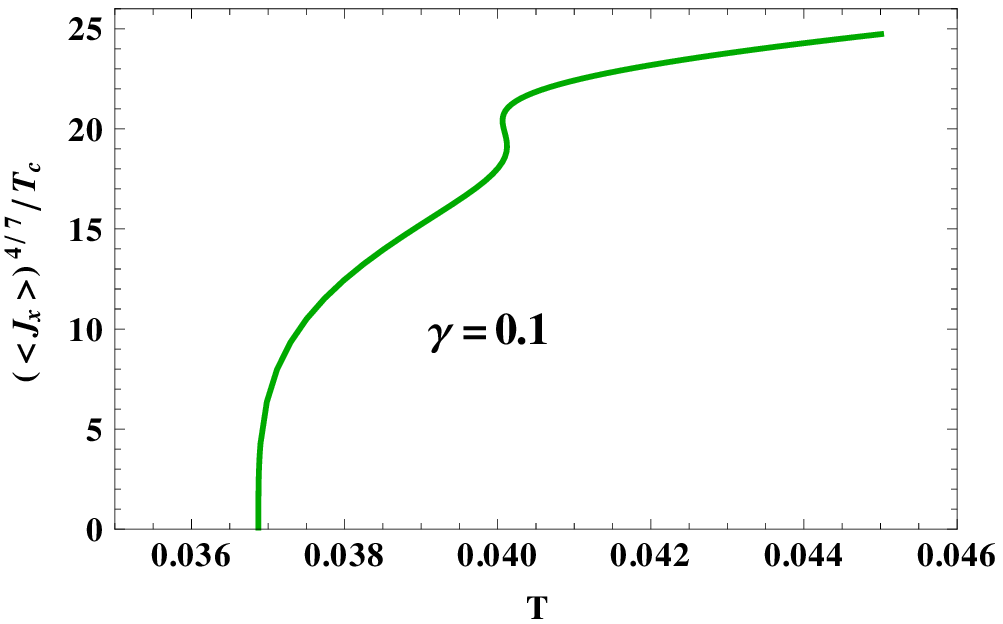}
\end{minipage}%
\caption{\label{fig:mass_3by16_condns6} Figures showing the condensate $(<J_x>)^{4/7}/T_c$ as a function of temperature for varying values of the Born-Infeld parameter $\gamma$ with the backreaction parameter fixed at $\kappa=1.01$. The blue, red and the green curves represent the different values of Born-Infeld parameter $\gamma=0.001,0.05$ and $0.1$ respectively. Here the mass of the vector field $\rho_x$ taken to be $m^2=-3/16$. } \end{figure} 

Subsequently we now study the effect of changing the Born-Infeld parameter on the thermodynamic behavior of the condensate. For this we consider the backreaction parameter to be $\kappa=0.8<\kappa_1$. In fig.(\ref{fig:mass_3by16_condns5}) we have plotted the condensate $(<J_x>)^{4/7}/T_c$ as a function of temperature for the value of the backreaction parameter, $\kappa=0.8<\kappa_1$ at different values of the Born-Infeld parameter, $\gamma=(0.001,0.05,0.1)$. From the plots it is observed that the condensate lies between two temperatures $T_0$ and $T_c$ describing a second order phase transition at $T_c$ and a zeroth order phase transition at $T_0$. For $\kappa<\kappa_1$, this behavior of the condensate remains the same for all values of the Born-Infeld parameter \footnote{We have checked the behavior of the condensate for higher values of $\gamma=1,10$ etc. and found out the behavior of the condensate doesn't changes for values of the backreaction parameter $\kappa<\kappa_1$}. Similarly in the fig.(\ref{fig:mass_3by16_condns6}),  we have plotted the condensate $(<J_x>)^{4/7}/T_c$ as a function of temperature for the value of the backreaction parameter, $\kappa=1.05>\kappa_2$ at different values of the Born-Infeld parameter, $\gamma=(0.001,0.05,0.1)$. For the case $\kappa>\kappa_2$, the condensate remains thermodynamically unfavorable for all values of $\gamma$ depicting a “ retrograde condensation ”. Here the interesting case arises when we take the value of the backreaction parameter $(\kappa=1.01)$ in between $\kappa_1$ and $\kappa_2$. In this case as we increase the value of $\gamma$ the vector field shows a “ retrograde condensation ” beyond a certain critical value of the Born-Infeld parameter $(\gamma^1_c=0.0254)$. Whereas below this critical value of $\gamma$ the condensate shows a first order phase transition at the critical temperature $T_c$ and a zeroth order phase transition at the temperature $T_0$.
\begin{figure}[H]
\centering
\begin{minipage}[b]{0.5\linewidth}
\includegraphics[width =2.5in,height=1.6in]{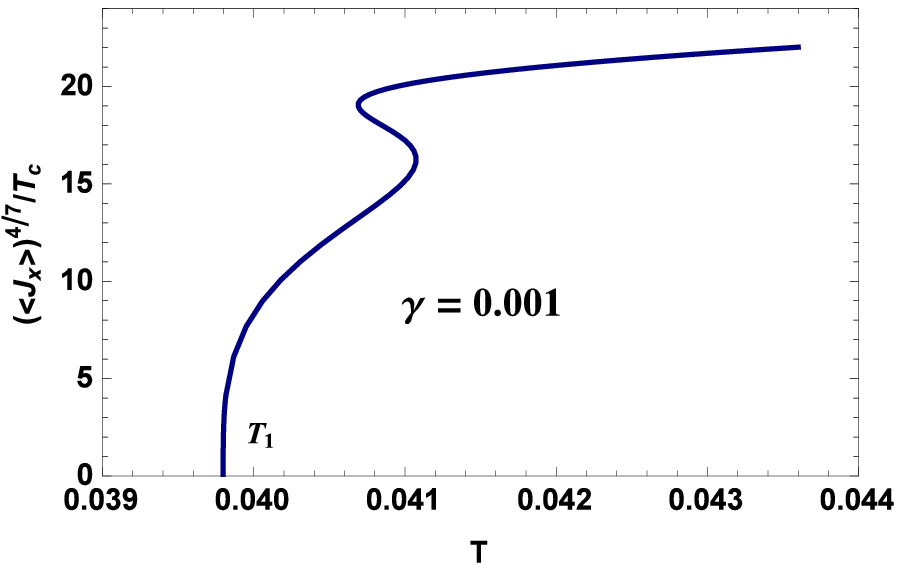}
\end{minipage}%
\begin{minipage}[b]{0.5\linewidth}
\includegraphics[width =2.5in,height=1.6in]{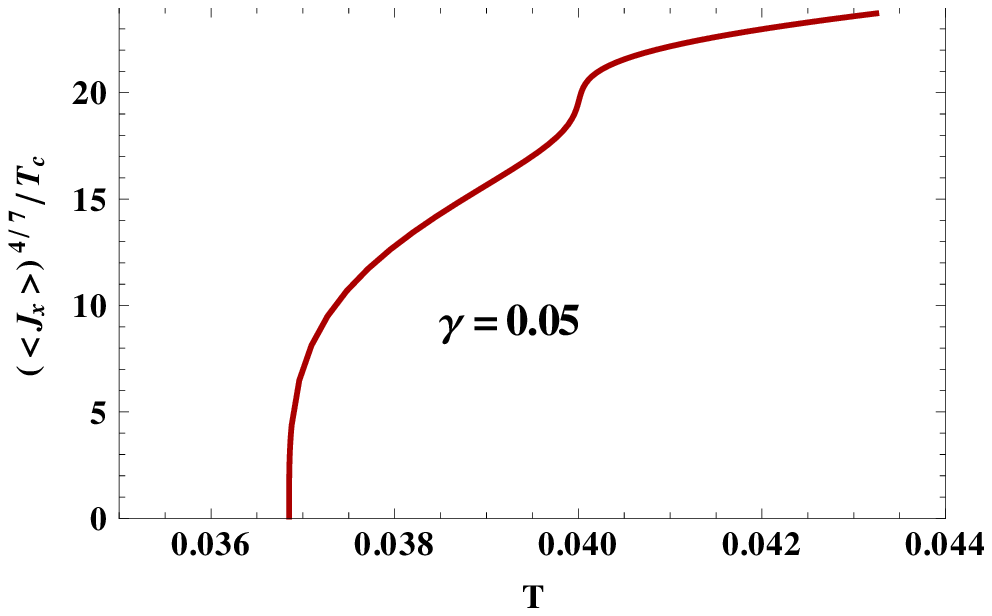}
\end{minipage}\quad
\begin{minipage}[b]{0.5\linewidth}
\includegraphics[width =2.5in,height=1.6in]{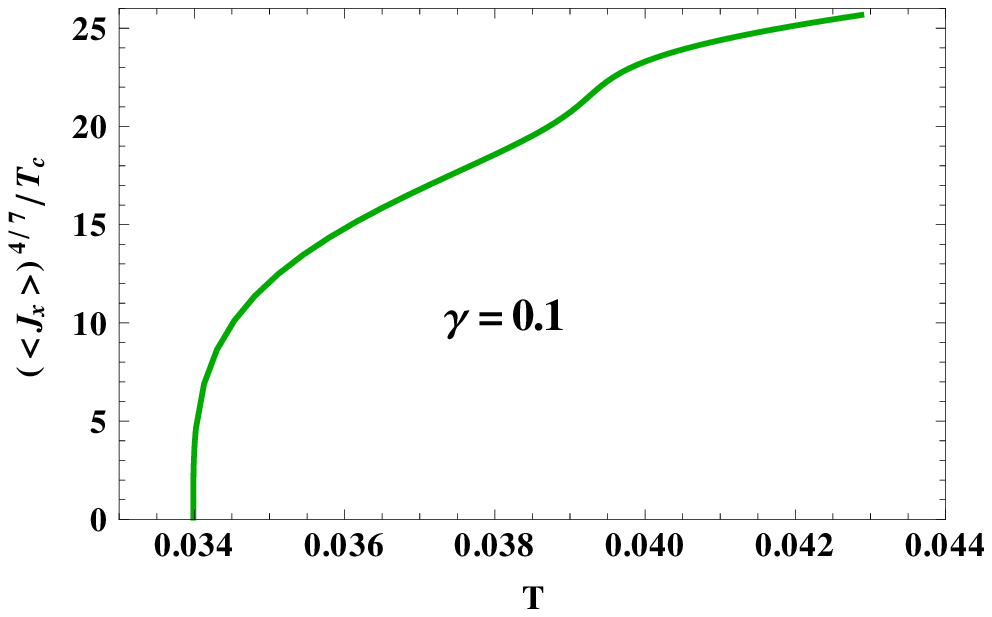}
\end{minipage}%
\caption{\label{fig:mass_3by16_condns7}  Figures showing the condensate $(<J_x>)^{4/7}/T_c$ as a function of temperature for varying values of the Born-Infeld parameter $\gamma$ with the backreaction parameter fixed at $\kappa=1.05$. The blue, red and the green curves represent the different values of Born-Infeld parameter $\gamma=0.001,0.05$ and $0.1$ respectively. Here the mass of the vector field $\rho_x$ taken to be $m^2=-3/16$. } \end{figure} 

We summarize the main results of this subsection by plotting the  $(T,\kappa)$ and $(T,\gamma)$ phase diagram in fig.(\ref{fig:Phase_mass_3by16}). The upper curve in the $(T,\kappa)$ phase diagram at fixed $\gamma=0.001$, represents the critical transition temperature $T_c$ for the regions $\kappa <\kappa_1$, $\kappa_1<\kappa<\kappa_2$ and $\kappa_2<\kappa$ respectively.  In contrast the lower curve represents the temperature $T_0$ corresponding to the zeroth order phase transition. At $\kappa_2$ the two curves merge to the single curve representing the {\it retrograde phase transition} temperature $T_1$. The regions below the $T_0$ curve and above the $T_c$ curve represent the normal phase of the boundary theory whereas the region in between these curves represents the condensed phase. The bounded green region existing below $\kappa_1$ represents the condensation via a second order phase transition and the bounded red region between $\kappa_1<\kappa<\kappa_2$ represents the condensation via a first order phase transition. Furthermore in the $(T,\gamma)$ diagram corresponding to $\kappa_1<\kappa=1.01<\kappa_c$, the upper curve represents the critical transition temperature $T_c$ whereas the lower curve represents the temperature $T_0$ corresponding to the zeroth order phase transition. At $\gamma^1_c$ the two curves merge to the single curve representing the {\it retrograde phase transition temperature} $T_1$. The regions below the $T_0$ curve and above the $T_c$ curve represent the normal phase of the boundary theory whereas the region in between these curves represents the condensed phase. The bounded red region existing below $\gamma^1_c$ represents the condensation via a first order phase transition. The phase diagrams also shows that as we increase $\kappa$ or $\gamma$ the critical temperature $T_c$ and the zeroth order phase transition temperature $T_0$ decrease gradually which  implies that the superconducting phase transition is hindered for large values of the backreaction parameter $\kappa$ or the Born-Infeld parameter $\gamma$ .

\section{The AC Conductivity}
In this section we focus on studying transport phenomena for the holographic superconductor, that describes the linear response of the system to small external sources. In particular we obtain the ac conductivity as a function of frequency for the p-wave superconducting phase of the strongly coupled (2+1)-dimensional boundary field theory. For the linear response framework considered by us this involves the computation of the retarded Green function that describes the current-current correlation in the boundary field theory. By using the GKPW prescription \cite{Gubser:1998bc,Witten:1998qj}, these correlation functions may be computed by studying the linear response\footnote{ It is to be noted that we are working in the the long wavelength and low frequency limit which is in relation to the linear response theory.} of the boundary theory to the fluctuations of the bulk gauge field $A_\mu$. These fluctuations are dual to the boundary electric current (say ${\cal J}$) such that $<{\cal J}_i>=G_{ij}^{ret}{\cal A}^j$, where $G_{ij}^{ret}$ stands for the retarded Green function and ${\cal A}_j$ stand for the external vector potential. For the Ohm's law we also have $<{\cal J}_i>=\sigma_{ij} {\cal E}^j$, where $\sigma_{ij}$ represents the electrical conductivity and ${\cal E}_j$ stands for the external electric field. Now from the aforementioned relations  the electrical conductivity may be deduced as, 
$\sigma_{ij}=G_{ij}^{ret}{\cal A}_k {\cal E}^k/{\cal E}^2$. In order to compute the ac conductivity for the condensed phase of the boundary theory one has to add the vector perturbation $e^{−i\omega t}A_\mu$ and the metric perturbation $e^{−i\omega t}g_{\mu t}$ to the hairy black hole background and then solve the linearized equations obtained from the Maxwell equation (\ref{MaxEom}) and the Einstein equation (\ref{EinsEom}).

As described above, the evaluation of the conductivity in the $x$-direction requires the addition of a vector perturbation $e^{−i\omega t}A_x(r)$ and a metric perturbation $e^{−i\omega t}g_{xt}(r)$ to the bulk configuration.  In this case the metric perturbation is coupled to the vector field $\rho_x$ even at linear order through the equation of motion (\ref{RhoEom}) for the vector field. This renders the numerical computation of  the ac conductivity 
$(\sigma_x)$  in the $x$ direction,  highly complicated and we omit this. However the calculation simplifies considerably for the ac conductivity $(\sigma_y)$ in the $y$-direction as the metric perturbation do not couple to the vector field $\rho_x$. For this case, we perturb the bulk by adding the vector  perturbation $e^{−i\omega t}A_y(r)$ and the metric perturbation $e^{−i\omega t}g_{yt}(r)$. These perturbations for the metric and the gauge field lead to two coupled equations in terms of $g_{yt}$ and $A_y$. Eliminating $g_{yt}$ from the coupled equations of motion we obtain a linearized equation for the gauge field $A_y$, which may be given as

\begin{eqnarray}
A_y''&+& A_y' \left(\frac{f'}{f}+\frac{h'}{h}+\frac{\chi ' \left(2 \gamma  e^{\chi} \phi'^2-1\right)+2\gamma  e^{\chi} \phi' \phi''}{2 \left(1-\gamma  e^{\chi} \phi'^2\right)}\right)\nonumber\\
&+& A_y\left(\frac{\omega ^2 e^{\chi}}{f^2}+\frac{2 \rho_x^2 \left(\gamma  e^{\chi} \phi'^2-1\right)-\kappa ^2 r^2 h e^{\chi} \phi'^2}{r^2 f h \sqrt{1-\gamma  e^{\chi} \phi'^2}}\right)=0.\label{Ayeom}
\end{eqnarray}

An analytic solution for the above differential equation seems computationally intractable, thus we proceed to solve it numerically. For this we begin by considering the horizon $(r=r_h=1)$ boundary conditions for $A_y(r)$ which may be written down as

\begin{eqnarray}
A_y(r)=f(r)^\delta S(r),~~ S(r)=(1+ a (1- r)+b(1- r)^2+\cdots),\label{hanastz}
\end{eqnarray}
where the coefficients $a$ and $b$ are functions of $\mu, \kappa,\gamma$ and $\omega$ and the dots represent the terms in higher powers of $(1-r)$. Now substituting  $A_y(r)=f(r)^\delta S(r)$ in eq.(\ref{Ayeom}) and taking the near horizon limit $(r \rightarrow 1)$ one may obtain the following expression for the exponent $\delta$,

\begin{eqnarray}
\delta=\pm \left.\frac{i e^{\frac{\chi}{2}}\omega}{f'}\right|_{r=1}
\end{eqnarray}

Taking $\delta = \delta_{-}~$, the near horizon form of $A_y(r)$ with the incoming wave boundary condition may be expressed as,

\begin{eqnarray}
A_y(r)=f(r)^{-\frac{i\omega e^{\chi(1)/2}}{f'(1)}} (1+ a(1- r)+b(1- r)^2+\cdots)
\end{eqnarray}

Similarly the asymptotic large $ r$ behavior of the perturbation $A_y(r)$ may be written down as
\begin{eqnarray}
A_y(r)=A^{(0)}_y+\frac{A^{(1)}_y}{r}+\cdots
\end{eqnarray}
here the leading term $A^{(0)}_y$ determines the source whereas the `normalisable' term $A^{(1)}_y$ gives the expectation value for the current in the dual boundary field theory respectively . Thus from the AdS/CFT dictionary we obtain 
\begin{eqnarray}
{\cal A}_y=A^{(0)}_y,~~<{\cal J}_x>=A^{(1)}_y,~~{\cal E}_x =\partial_t A^{(0)}_y=i\omega A^{(0)}_y
\end{eqnarray}

Now from Ohm's law, the expression for the ac conductivity in the $y$-direction may be expressed as follows

\begin{eqnarray}
\sigma_y(\omega,\kappa,\gamma)=\frac{<{\cal J}_x>}{{\cal E}_x}=-\frac{i A^{(1)}_y}{\omega A^{(0)}_y}.\label{sigmaexpr}
\end{eqnarray}

Using the expression given in equation (\ref{sigmaexpr}),  we numerically compute the ac conductivity for the superconducting phase of the (2+1)- dimensional boundary field theory.
\begin{figure}[H]
\centering
\begin{minipage}[b]{0.5\linewidth}
\includegraphics[width =2.5in,height=1.6in]{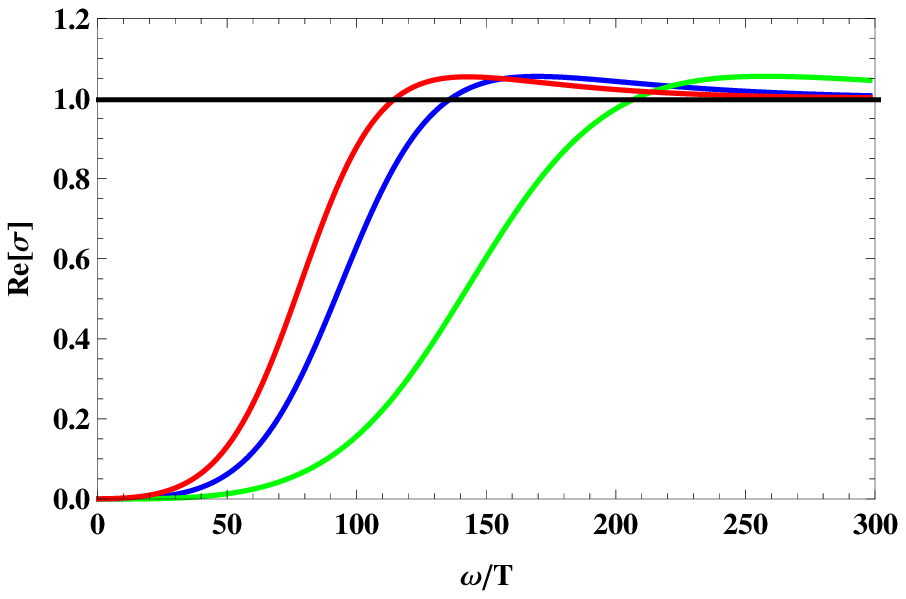}
\end{minipage}%
\begin{minipage}[b]{0.5\linewidth}
\includegraphics[width =2.5in,height=1.6in]{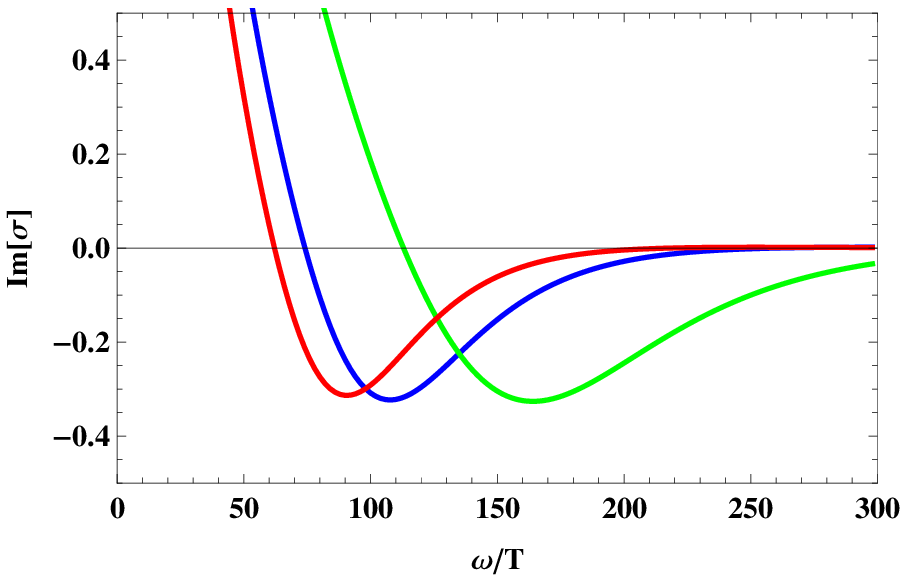}
\end{minipage}%
\caption{\label{fig:sigmakgFix3by4} For $m^2=3/4$, real and imaginary parts of the ac conductivity for superconducting phase are plotted with respect to $\omega/T$ for fixed values of $\kappa=0.7$ and $\gamma=0.05$. The red, blue and green curves correspond to the temperatures $T/T_c=0.8753$ , $T/T_c=0.5743$ and $T/T_c=1.0404$ respectively. } \end{figure} 
In fig.(\ref{fig:sigmakgFix3by4}),  we have plotted the real and imaginary parts of the ac conductivity  as a function of $\omega/T$ which correspond to the condensate for $m^2=3/4, \kappa=0.7$ and $\gamma=0.05$ displayed in fig.(\ref{fig:mass_3by4_condns3}). The horizontal line
in the plots for the real part of the ac conductivity corresponds  to the temperatures at or above the critical value for which there is no condensate. This line also represents the frequency independent conductivity for the normal phase of the boundary theory \cite{Hartnoll:2008vx}. From the figure it may be observed that for the superconducting phase, the real part of the ac conductivity shows a gapped behavior for temperatures below $T_c$. This gap in the real part of the ac conductivity disappears for very small values of the frequency ($\omega$)  whereas,  for higher values of $\omega$ the conductivity approaches asymptotically to the frequency independent conductivity for the normal phase. Furthermore, for temperatures below than the critical temperature  (i.e. $T<Tc$) there is also a delta function at $\omega=0$, which corresponds to a pole in the imaginary part of the ac conductivity such that the Kramers-Kronig relations are satisfied \cite{Hartnoll:2008vx}.

\begin{figure}[H]
\centering
\begin{minipage}[b]{0.5\linewidth}
\includegraphics[width =2.5in,height=1.6in]{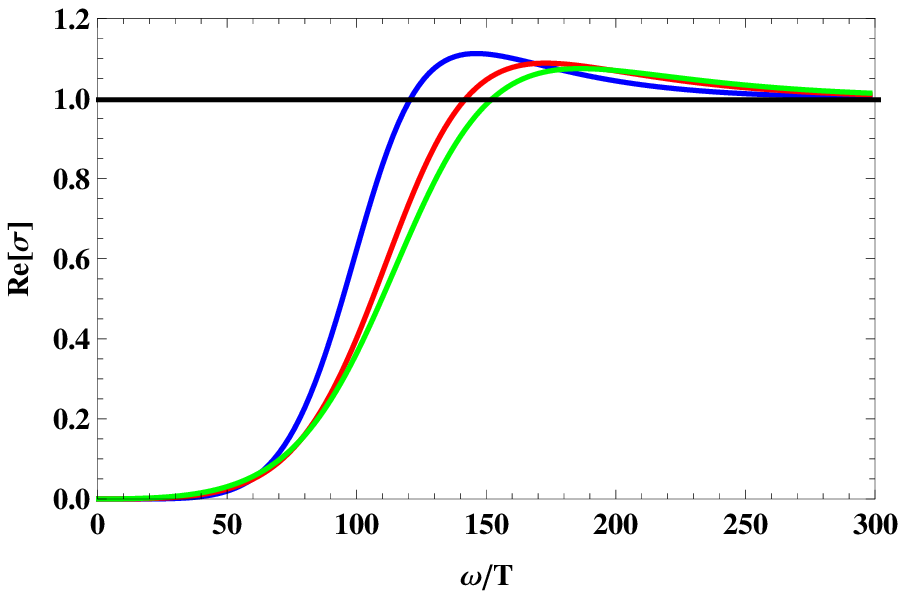}
\end{minipage}%
\begin{minipage}[b]{0.5\linewidth}
\includegraphics[width =2.5in,height=1.6in]{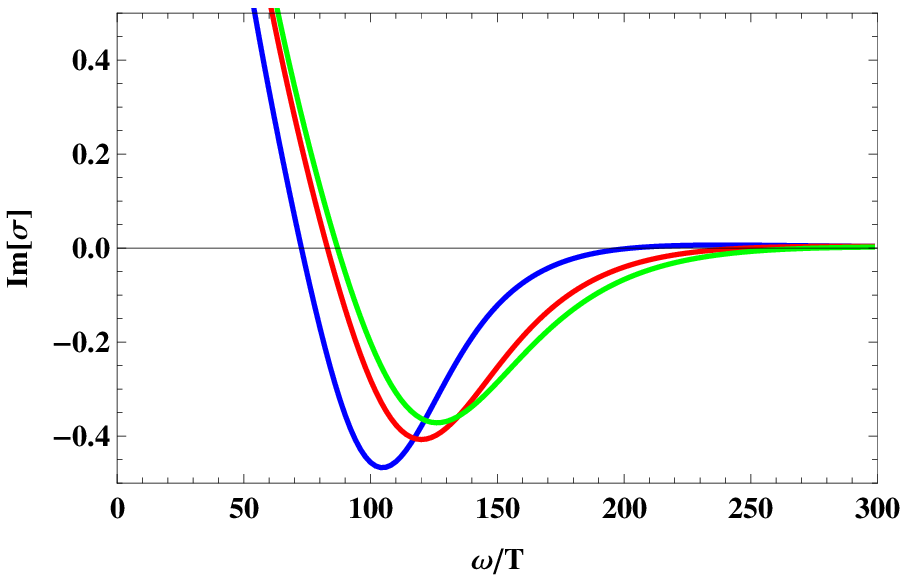}
\end{minipage}%
\caption{\label{fig:sigmagFix_k_3by4} For $m^2=3/4$, real and imaginary parts of the ac conductivity for superconducting phase are plotted with respect to $\omega/T$ for fixed values of $\gamma=0.001$ and temperature $T=0.0038\mu$. The blue, red and the green curves correspond to the values of the backreaction parameter  $\kappa=0.5$ , $\kappa=0.7$ and $\kappa=0.8$ respectively.} \end{figure} 
\begin{figure}[H]
\centering
\begin{minipage}[b]{0.5\linewidth}
\includegraphics[width =2.5in,height=1.6in]{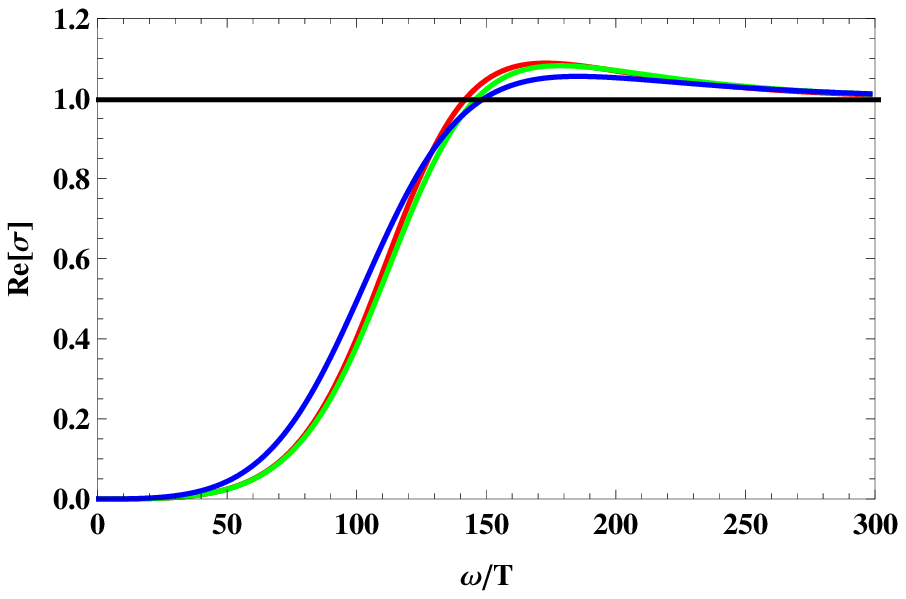}
\end{minipage}%
\begin{minipage}[b]{0.5\linewidth}
\includegraphics[width =2.5in,height=1.6in]{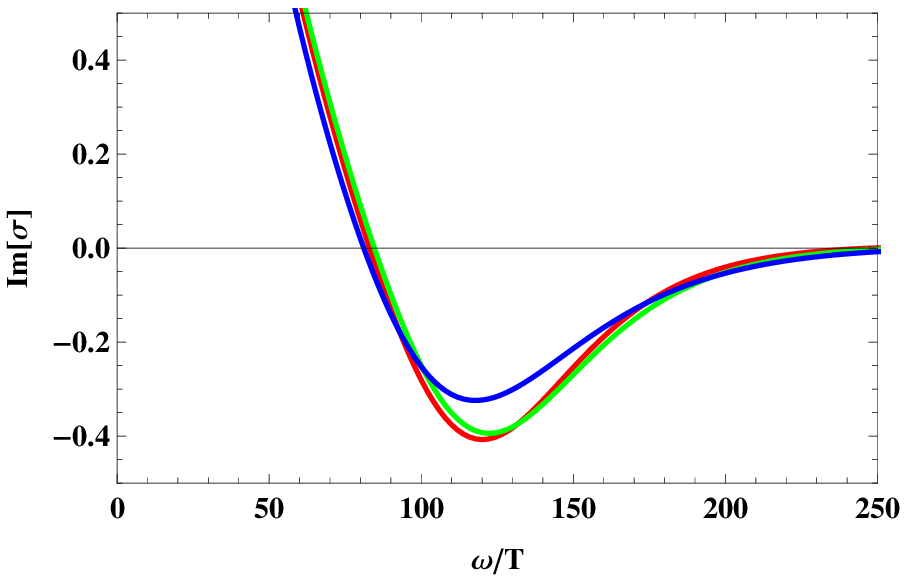}
\end{minipage}%
\caption{\label{fig:sigmakFix_g_3by4}  For $m^2=3/4$, real and imaginary parts of the ac conductivity for superconducting phase are plotted with respect to $\omega/T$ for fixed values of $\kappa=0.7$ and the temperature $T=0.0038 \mu$. The red, green and the blue curves correspond to the values of the backreaction parameter  $\gamma=0.001$ , $\gamma=0.008$ and $\gamma=0.05$ respectively. } \end{figure} 
As described earlier, we have explored the vector condensate for two different values of the mass of the vector field lying in different regimes ($m^2=-3/16<m^2_c$ and $m^2=3/4>m^2_c$). We now describe in detail the behaviour of the ac conductivity for these two different values of the mass for the vector field. For $m^2=3/4$, in fig.(\ref{fig:sigmagFix_k_3by4}) we have plotted the real and imaginary parts of the ac conductivity with $\omega/T$ with different values of the backreaction parameter for fixed value of $\gamma=0.001$ and the temperature $T=0.0038 \mu$. It may be observed from the figure that the gap in the real part of ac conductivity shifts to higher values of the frequency $(\omega)$ for increasing values of the backreaction parameter. Furthermore, the conductivity peaks decrease for  higher values of the backreaction parameter $\kappa$. Similarly in fig.(\ref{fig:sigmakFix_g_3by4}), we have plotted the real and imaginary parts of the ac conductivity with $\omega/T$ with different values of the Born-Infeld parameter $\gamma$ for fixed value of $\kappa=0.7$ and the temperature $T=0.0038\mu$. From the figure it may be observed that for increasing values of the Born-Infeld parameter, the gap in the real part of ac conductivity shifts to higher values of the frequency $(\omega)$ with the decrease in the conductivity peaks. 
\begin{figure}[H]
\centering
\begin{minipage}[b]{0.5\linewidth}
\includegraphics[width =2.5in,height=1.6in]{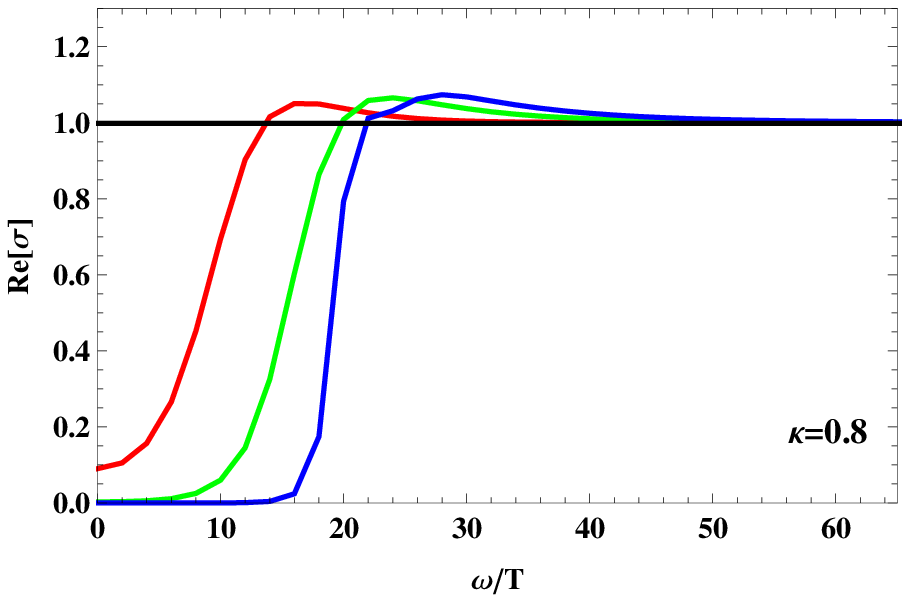}
\end{minipage}%
\begin{minipage}[b]{0.5\linewidth}
\includegraphics[width =2.5in,height=1.6in]{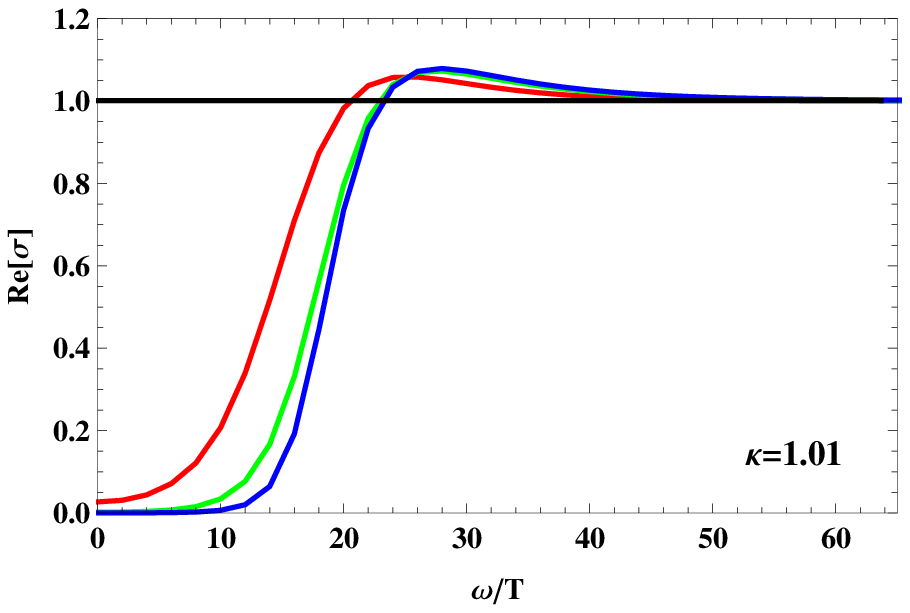}
\end{minipage}\quad
\begin{minipage}[b]{0.5\linewidth}
\includegraphics[width =2.5in,height=1.6in]{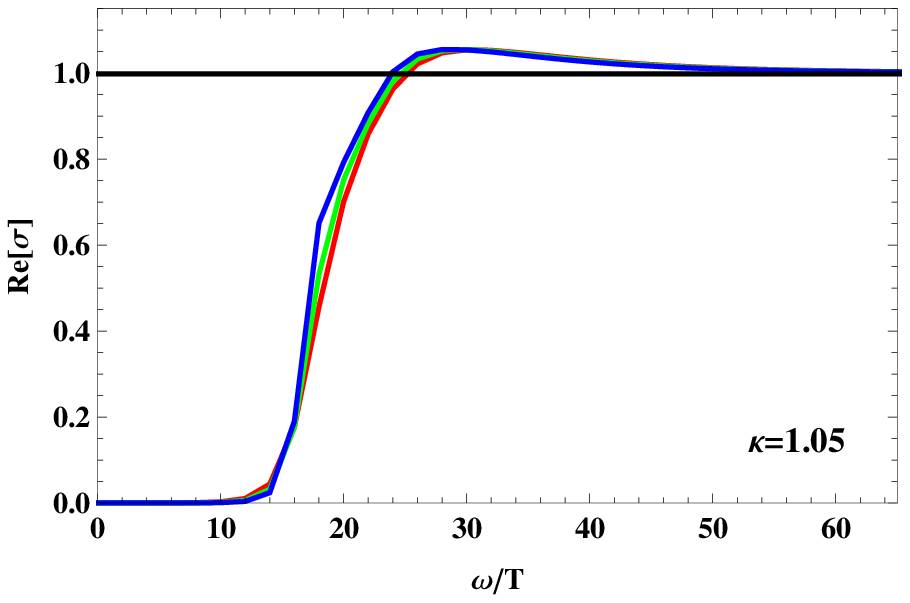}
\end{minipage}%
\begin{minipage}[b]{0.5\linewidth}
\includegraphics[width =2.5in,height=1.6in]{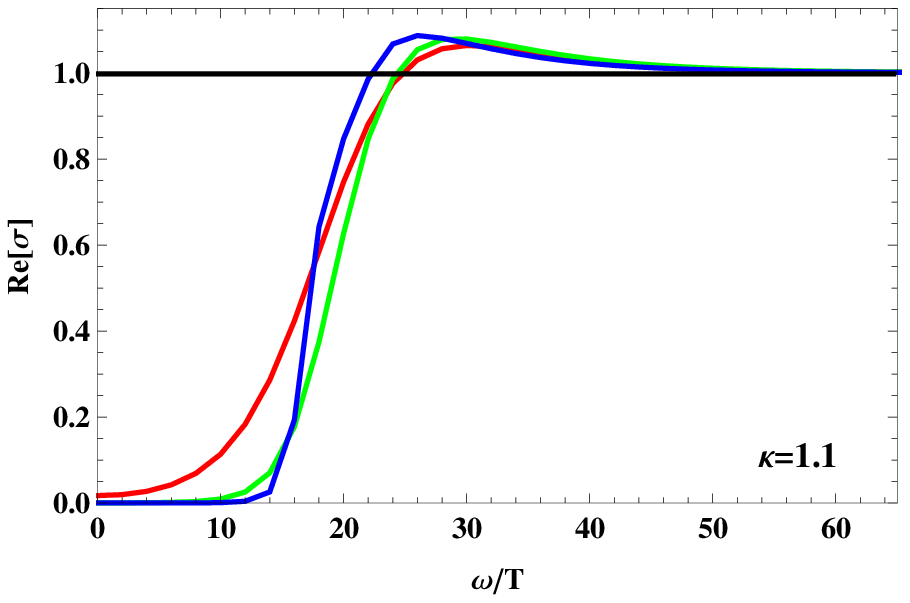}
\end{minipage}%
\caption{\label{fig:sigmak3by16_gFix1}  For $m^2=-3/16$, real and imaginary parts of the ac conductivity for superconducting phase are plotted with respect to $\omega/T$ for fixed value of $\gamma=0.001$ and varying values of $\kappa$. For $\kappa=0.8,1.01$, the red, green and blue curves in each graph correspond to decreasing value of the temperature. Whereas for $\kappa=1.05,1.1$, the red, green and blue curves in each graph correspond to increasing value of the temperature.} \end{figure} 
\begin{figure}[H]
\centering
\begin{minipage}[b]{0.5\linewidth}
\includegraphics[width =2.5in,height=1.6in]{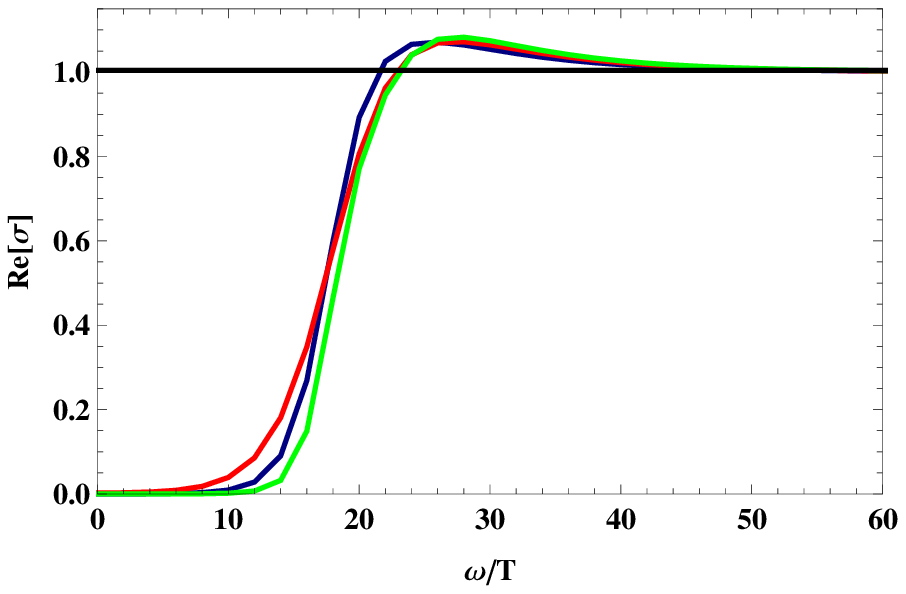}
\end{minipage}%
\begin{minipage}[b]{0.5\linewidth}
\includegraphics[width =2.5in,height=1.6in]{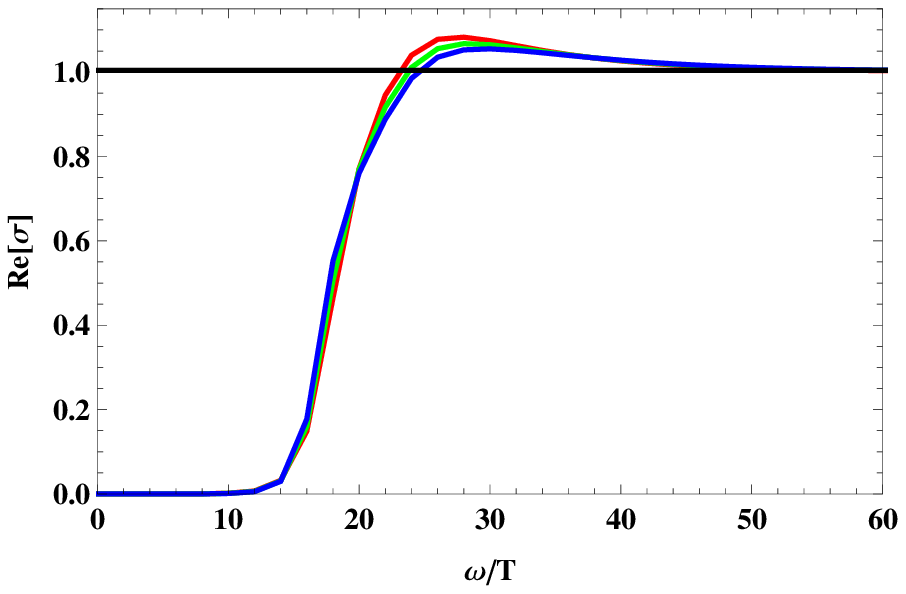}
\end{minipage}%
\caption{\label{fig:sigmak3by16}  For $m^2=-3/16$, real and imaginary parts of the ac conductivity for superconducting phase are plotted with respect to $\omega/T$. The left graph is for a fixed value of $\gamma=0.001$ and $T=0.0419\mu$ with red, green and the blue curves corresponding to $\kappa=0.8$ , $\kappa=1.01$ and $\kappa=1.05$ respectively. Whereas the right graph is for a fixed value of $\kappa=1.05$ and $T=0.0419\mu$ with red, green and the blue curves corresponding to $\gamma=0.001$ , $\gamma=0.05$ and $\gamma=0.1$ respectively.} \end{figure} 

For the mass $m^2=-3/16$, in fig.(\ref{fig:sigmak3by16_gFix1}) we have plotted the real and imaginary parts of the ac conductivity for different values of the backreaction parameter $\kappa$ and fixed value of Born-Infeld parameter $(\gamma=0.001)$ at different temperatures. In this case the parameter space of $\kappa$ is partitioned into three regions $\kappa <\kappa_1$, $\kappa_1<\kappa<\kappa_2$ and $\kappa_2<\kappa$ respectively and the behavior of the condensate changes radically in these regions. For the values of $\kappa=(0.8,1.01)$ less than $\kappa_2$, we observe the opening of a gap in the real part of the ac conductivity  the temperature is lowered below the critical temperature. But for the values of $\kappa=(1.05,1.1)$ greater than $\kappa_2$ , the opening of a gap in real part of the ac conductivity occurs for temperatures above than the critical temperature. This may be related to the fact that for this case the condensate formation  occurs via a {\it retrograde phase transition}  and the condensate formation occurs only for temperatures above the critical temperature. Furthermore, in the left graph of fig.(\ref{fig:sigmak3by16}) we have plotted the real and imaginary parts of the ac conductivity with $\omega/T$ for different values of the backreaction parameter and fixed value of $\gamma=0.001$ at temperature $T=0.0419 \mu$. It may be observed from the figure that the gap in the real part of the ac conductivity shifts to higher values of the frequency $(\omega)$ for  increasing  values of the backreaction parameter. The conductivity peaks also decrease for higher values of the backreaction parameter $\kappa$. Similarly,  in the right graph of the fig.(\ref{fig:sigmak3by16}) we have plotted the real and imaginary parts of the ac conductivity with $\omega/T$ for different values of the Born-Infeld parameter and fixed value of $\kappa=1.05$ at temperature $T=0.0419\mu$. From the figure it may also be observed that foe increasing values of the Born-Infeld parameter, the gap in the real part of the ac conductivity shifts to higher values of the frequency $(\omega)$ with the decrease in  the conductivity peaks. 

\section{Summary and Conclusions}

In summary we have constructed a model of (2+1) dimensional p-wave holographic superconductors from charged Born-Infeld black holes in a $AdS_4$ bulk in the presence of a charged massive vector field. The superconducting phase of the strongly coupled boundary field theory corresponds to the condensation of the complex vector field leading to the formation of {\it vector hair} for the charged Born-Infeld black hole. For the boundary field theory this leads to a charged vector operator acquiring a vacuum expectation value that breaks both the $U(1)$ symmetry as well as the rotational invariance spontaneously. This coressponds to an anisotropic condensate in the boundary field theory that describes a p-wave holographic superconductor. Considering the backreaction of the vector matter fields we have elucidated a rich and varied phase structure for the boundary field theory depending on the relative values of the mass parameter  $m$, the backreaction parameter $\kappa$ and the Born-Infeld parameter $\gamma$. It has been shown in our construction that depending on the values of the parameters described above it is possible to obtain  zeroth order, first order, second order and retrograde phase transitions for the condensate formation in the strongly coupled (2+1) dimensional boundary field theory. While the phase structure obtained by us is similar to that described in  \cite{Cai:2013aca} in the context of a linear Maxwell electrodynamics, we also observe novel behaviours controlled by  the Born-Infeld parameter for our nonlinear scenario. From our construction we obtain a critical value ($m^2_c=0$) of the mass parameter $m^2$ such that the parameter space is partitioned into two regions with different behaviours of the condensate for different values of the parameters $\kappa$ and $\gamma$. Particularly it is observed that for  $m^2>m^2_c$ there is a critical value ($\kappa_c$)  for small values of $\gamma$ such that the condensate shows a second order phase transition for $\kappa<\kappa_c$ whereas the transition is of first order for $\kappa>\kappa_c$.
 
For  $m^2<m^2_c$, we have shown that at very small values of the Born-Infeld parameter $\gamma$ the system has two critical values of $\kappa$ namely, $\kappa_1$ and $\kappa_2$. For $\kappa<\kappa_1$ the condensate formation follows a second order phase transition whereas for $\kappa_1<\kappa<\kappa_2$, it is a first order phase transition. Finally for $\kappa_2<\kappa$, the condensation of the vector field occurs via a {\it retrograde phase transition} that exhibits a condensate formation for temperatures higher than the critical temperature $T_1$. This retrograde phase transition corresponds to a metastable phase of the condensate for which the free energy is higher than that of the normal phase. Furthermore we have shown from our construction that for fixed values of $\kappa$ for the different regions mentioned above between the two critical values $\kappa_1$ and $\kappa_2$, changing the Born-Infeld parameter 
$\gamma$ also affects the nature of the phase transition. Particularly for $\kappa<\kappa_1$ and $\kappa_2<\kappa$, changing $\gamma$ has no effect on the order of the phase transition whereas for $\kappa_1<\kappa<\kappa_2$, increasing $\gamma$ beyond a certain critical value $\gamma^1_c$ changes the phase transition from  first order  to a {\it retrograde phase transition}. The general behaviour which emerges from our analysis is that, irrespective of any value of $m^2$ the critical transition temperature  decreases with increasing values of the parameters $\kappa$ and $\gamma$. 

In order to characterize the nature of the different types of superconducting phase transitions in our model, we have studied the holographic free energy for both the normal and the superconducting phases for  different possible values of the parameters $m^2$, $\kappa$ and $\gamma$. For the first order phase transition the free energy shows a characteristic {\it swallow tail} behaviour whereas for the second order phase transition the free energy exhibits a kink at the critical temperature. Furthermore, for the zeroth order phase transition the free energy describes an abrupt discontinuity at the transition from the superconducting to the normal phase of the boundary field theory.  It is to be noted that the holographic free energy considered here may be considered as a generalized version of the usual Landau-Ginzburg free energy \cite{Aprile:2010yb} where the coefficient of the quadratic term in the order parameter depends linearly on the temperature whereas the coefficient of the fourth order term is weakly dependent. Also the holographic free energy considered here describes a strongly coupled phase of the boundary field theory which shows a deviation from the usual mean field behaviour.  In the context of the phenomenological Landau-Ginzburg theory such a deviation would imply an unusual dependence of the critical temperature on the higher order terms. Thus in this context an analytical description of our construction may serve to provide a deeper insight into this interesting issue. 

In contrast to \cite{Cai:2013aca} in a linear Maxwell scenario we have further computed the ac conductivity for the p-wave superconducting phase of the strongly coupled 2+1 dimensional boundary field theory. We observe from our construction that the ac conductivity exhibits the usual behaviour as obtained for the case of p-wave holographic superconductors described in \cite{Gubser:2008wv}. The only exception  occurs for the case of the condensate corresponding to the {\it retrograde phase transition} where the real part of the ac conductivity shows the  formation of a gap for temperatures above the critical temperature $T_1$. Finally in passing we would like to mention that the {\it retrograde phase transition} for the p-wave superconducting phase of the boundary field theory is usually observed for  binary and multicomponent liquid mixtures as described in \cite{Narayanan1994135}. This interesting type of phase transition also occurs in some of the superconducting materials such as the granular $BaPb_{0.75}Bi_{0.25}O_3$ compound \cite{PhysRevB.29.1493} and the cuprate superconductors described in \cite{PhysRevB.51.3134}. This naturally makes  it interesting to explore whether our construction would be relevant for the description of the phase structure and the transport properties of such superconducting materials.

Our work leads to extremely interesting future directions for investigations. It would be interesting to generalize our construction to the gravitational background lattice described in \cite{Horowitz:2012gst,Horowitz:2012gs}. Such an investigation  may lead to interesting behaviour of the transport properties such as the ac conductivity for such models. Furthermore it may also be possible to extend our construction to the case of the zero temperature gravitational background, such as the AdS soliton where the system may undergo a superconductor/insulator phase transition as described in \cite{Nishioka:2009zj,Bai:2012cx,Cai:2013oma,Chaturvedi:2014dga}. Additionally our construction may be generalized to study the phenomena of holographic quantum quench \cite{Das:2011nk,Basu:2011ft,Basu:2012gg} that may lead to interesting insights due to the presence of the non-linear Born-Infeld term considered by us. We leave these interesting avenues for future investigations.

\section{Acknowledgment}
The authors would like to thank Sayantani Bhattacharya, Supriya Kar and Kaushik Ray for useful discussions.  The work of Pankaj Chaturvedi is supported by the Grant No. 09/092(0846)/2012-EMR-I, from the Council of Scientific and Industrial Research (CSIR), India.
\bibliographystyle{unsrt}
\bibliography{mybib}

\begin{thebibliography}{10}

\bibitem{Maldacena:1997re}
Juan~Martin Maldacena.
\newblock {The Large N limit of superconformal field theories and
  supergravity}.
\newblock {\em Int.J.Theor.Phys.}, 38:1113--1133, 1999.

\bibitem{Gubser:1998bc}
S.S. Gubser, Igor~R. Klebanov, and Alexander~M. Polyakov.
\newblock {Gauge theory correlators from noncritical string theory}.
\newblock {\em Phys.Lett.}, B428:105--114, 1998.

\bibitem{Witten:1998qj}
Edward Witten.
\newblock {Anti-de Sitter space and holography}.
\newblock {\em Adv.Theor.Math.Phys.}, 2:253--291, 1998.

\bibitem{Aharony:1999ti}
Ofer Aharony, Steven~S. Gubser, Juan~Martin Maldacena, Hirosi Ooguri, and Yaron
  Oz.
\newblock {Large N field theories, string theory and gravity}.
\newblock {\em Phys.Rept.}, 323:183--386, 2000.

\bibitem{Witten:1998zw}
Edward Witten.
\newblock {Anti-de Sitter space, thermal phase transition, and confinement in
  gauge theories}.
\newblock {\em Adv.Theor.Math.Phys.}, 2:505--532, 1998.

\bibitem{Gubser:2008px}
Steven~S. Gubser.
\newblock {Breaking an Abelian gauge symmetry near a black hole horizon}.
\newblock {\em Phys.Rev.}, D78:065034, 2008.

\bibitem{Hartnoll:2008vx}
Sean~A. Hartnoll, Christopher~P. Herzog, and Gary~T. Horowitz.
\newblock {Building a Holographic Superconductor}.
\newblock {\em Phys.Rev.Lett.}, 101:031601, 2008.

\bibitem{Hartnoll:2008kx}
Sean~A. Hartnoll, Christopher~P. Herzog, and Gary~T. Horowitz.
\newblock {Holographic Superconductors}.
\newblock {\em JHEP}, 0812:015, 2008.

\bibitem{Herzog:2009xv}
Christopher~P. Herzog.
\newblock {Lectures on Holographic Superfluidity and Superconductivity}.
\newblock {\em J.Phys.}, A42:343001, 2009.

\bibitem{Horowitz:2010gk}
Gary~T. Horowitz.
\newblock {Introduction to Holographic Superconductors}.
\newblock {\em Lect.Notes Phys.}, 828:313--347, 2011.

\bibitem{Herzog:2010vz}
Christopher~P. Herzog.
\newblock {An Analytic Holographic Superconductor}.
\newblock {\em Phys.Rev.}, D81:126009, 2010.

\bibitem{Sachdev:2010ch}
Subir Sachdev.
\newblock {Condensed Matter and AdS/CFT}.
\newblock {\em Lect.Notes Phys.}, 828:273--311, 2011.

\bibitem{Roberts:2008ns}
Matthew~M. Roberts and Sean~A. Hartnoll.
\newblock {Pseudogap and time reversal breaking in a holographic
  superconductor}.
\newblock {\em JHEP}, 0808:035, 2008.

\bibitem{Horowitz:2009ij}
Gary~T. Horowitz and Matthew~M. Roberts.
\newblock {Zero Temperature Limit of Holographic Superconductors}.
\newblock {\em JHEP}, 0911:015, 2009.

\bibitem{Kim:2013oba}
Keun-Young Kim and Marika Taylor.
\newblock {Holographic d-wave superconductors}.
\newblock {\em JHEP}, 1308:112, 2013.

\bibitem{Benini:2010pr}
Francesco Benini, Christopher~P. Herzog, Rakibur Rahman, and Amos Yarom.
\newblock {Gauge gravity duality for d-wave superconductors: prospects and
  challenges}.
\newblock {\em JHEP}, 1011:137, 2010.

\bibitem{Chen:2010mk}
Jiunn-Wei Chen, Ying-Jer Kao, Debaprasad Maity, Wen-Yu Wen, and Chen-Pin Yeh.
\newblock {Towards A Holographic Model of D-Wave Superconductors}.
\newblock {\em Phys.Rev.}, D81:106008, 2010.

\bibitem{Gubser:2008wv}
Steven~S. Gubser and Silviu~S. Pufu.
\newblock {The Gravity dual of a p-wave superconductor}.
\newblock {\em JHEP}, 0811:033, 2008.

\bibitem{Ammon:2009xh}
Martin Ammon, Johanna Erdmenger, Viviane Grass, Patrick Kerner, and Andy
  O'Bannon.
\newblock {On Holographic p-wave Superfluids with Back-reaction}.
\newblock {\em Phys.Lett.}, B686:192--198, 2010.

\bibitem{Zeng:2010fs}
Hua-Bi Zeng, Wei-Min Sun, and Hong-Shi Zong.
\newblock {Supercurrent in p-wave Holographic Superconductor}.
\newblock {\em Phys.Rev.}, D83:046010, 2011.

\bibitem{Cai:2010zm}
Rong-Gen Cai, Zhang-Yu Nie, and Hai-Qing Zhang.
\newblock {Holographic Phase Transitions of P-wave Superconductors in
  Gauss-Bonnet Gravity with Back-reaction}.
\newblock {\em Phys.Rev.}, D83:066013, 2011.

\bibitem{Aprile:2010ge}
Francesco Aprile, Diego Rodriguez-Gomez, and Jorge~G. Russo.
\newblock {p-wave Holographic Superconductors and five-dimensional gauged
  Supergravity}.
\newblock {\em JHEP}, 1101:056, 2011.

\bibitem{Zayas:2011dw}
Leopoldo~A. Pando~Zayas and Dori Reichmann.
\newblock {A Holographic Chiral $p_x + ip_y$ Superconductor}.
\newblock {\em Phys.Rev.}, D85:106012, 2012.

\bibitem{Momeni:2012ab}
D.~Momeni, N.~Majd, and R.~Myrzakulov.
\newblock {p-wave holographic superconductors with Weyl corrections}.
\newblock {\em Europhys.Lett.}, 97:61001, 2012.

\bibitem{Roychowdhury:2013aua}
Dibakar Roychowdhury.
\newblock {Holographic droplets in p-wave insulator/superconductor transition}.
\newblock {\em JHEP}, 1305:162, 2013.

\bibitem{Cai:2013oma}
Rong-Gen Cai, Li~Li, Li-Fang Li, and Ru-Keng Su.
\newblock {Entanglement Entropy in Holographic P-Wave Superconductor/Insulator
  Model}.
\newblock {\em JHEP}, 1306:063, 2013.

\bibitem{Cai:2013pda}
Rong-Gen Cai, Song He, Li~Li, and Li-Fang Li.
\newblock {A Holographic Study on Vector Condensate Induced by a Magnetic
  Field}.
\newblock {\em JHEP}, 1312:036, 2013.

\bibitem{Cai:2013aca}
Rong-Gen Cai, Li~Li, and Li-Fang Li.
\newblock {A Holographic P-wave Superconductor Model}.
\newblock {\em JHEP}, 1401:032, 2014.

\bibitem{Maslov:2004ap}
V.P. Maslov.
\newblock Zeroth-order phase transitions.
\newblock {\em Mathematical Notes}, 76(5-6):697--710, 2004.

\bibitem{Narayanan1994135}
T.~Narayanan and Anil Kumar.
\newblock Reentrant phase transitions in multicomponent liquid mixtures.
\newblock {\em Physics Reports}, 249(3):135 -- 218, 1994.

\bibitem{Horowitz:2010jq}
Gary~T. Horowitz and Benson Way.
\newblock {Complete Phase Diagrams for a Holographic Superconductor/Insulator
  System}.
\newblock {\em JHEP}, 1011:011, 2010.

\bibitem{Peng:2011gh}
Yan Peng, Qiyuan Pan, and Bin Wang.
\newblock {Various types of phase transitions in the AdS soliton background}.
\newblock {\em Phys.Lett.}, B699:383--387, 2011.

\bibitem{Cai:2012es}
Rong-Gen Cai, Song He, Li~Li, and Li-Fang Li.
\newblock {Entanglement Entropy and Wilson Loop in St\'{u}ckelberg Holographic
  Insulator/Superconductor Model}.
\newblock {\em JHEP}, 1210:107, 2012.

\bibitem{Horowitz:2008bn}
Gary~T. Horowitz and Matthew~M. Roberts.
\newblock {Holographic Superconductors with Various Condensates}.
\newblock {\em Phys.Rev.}, D78:126008, 2008.

\bibitem{Born:1934trs}
M.~Born and L.~Infeld.
\newblock Foundations of the new field theory.
\newblock {\em Proceedings of the Royal Society of London A: Mathematical,
  Physical and Engineering Sciences}, 144(852):425--451, 1934.

\bibitem{Gibbons1995185}
G.W. Gibbons and D.A. Rasheed.
\newblock Electric-magnetic duality rotations in non-linear electrodynamics.
\newblock {\em Nuclear Physics B}, 454(1-2):185--206, 1995.
\newblock cited By 217.

\bibitem{Fradkin1985123}
E.S. Fradkin and A.A. Tseytlin.
\newblock Non-linear electrodynamics from quantized strings.
\newblock {\em Physics Letters B}, 163(1-4):123--130, 1985.
\newblock cited By 521.

\bibitem{Dey:2004yt}
Tanay~Kr. Dey.
\newblock {Born-Infeld black holes in the presence of a cosmological constant}.
\newblock {\em Phys.Lett.}, B595:484--490, 2004.

\bibitem{Cai:2004eh}
Rong-Gen Cai, Da-Wei Pang, and Anzhong Wang.
\newblock {Born-Infeld black holes in (A)dS spaces}.
\newblock {\em Phys.Rev.}, D70:124034, 2004.

\bibitem{Jing201068}
Jiliang Jing and Songbai Chen.
\newblock Holographic superconductors in the born–infeld electrodynamics.
\newblock {\em Physics Letters B}, 686(1):68 -- 71, 2010.

\bibitem{Gangopadhyay:2013qza}
Sunandan Gangopadhyay.
\newblock {Holographic superconductors in Born-Infeld electrodynamics and
  external magnetic field}.
\newblock {\em Mod.Phys.Lett.}, A29:1450088, 2014.

\bibitem{Jing:2011sv}
Jiliang Jing, Qiyuan Pan, and Songbai Chen.
\newblock Holographic superconductors with power-maxwell field.
\newblock {\em Journal of High Energy Physics}, 2011(11), 2011.

\bibitem{PhysRevD.83.066010}
Jiliang Jing, Liancheng Wang, Qiyuan Pan, and Songbai Chen.
\newblock Holographic superconductors in gauss-bonnet gravity with born-infeld
  electrodynamics.
\newblock {\em Phys. Rev. D}, 83:066010, Mar 2011.

\bibitem{Gangopadhyay:2012sv}
Sunandan Gangopadhyay and Dibakar Roychowdhury.
\newblock Analytic study of properties of holographic superconductors in
  born-infeld electrodynamics.
\newblock {\em Journal of High Energy Physics}, 2012(5), 2012.

\bibitem{Liu:2012hc}
Yunqi Liu, Yan Peng, and Bin Wang.
\newblock {Gauss-Bonnet holographic superconductors in Born-Infeld
  electrodynamics with backreactions}.
\newblock 2012.

\bibitem{PhysRevD.86.106009}
Dibakar Roychowdhury.
\newblock Effect of external magnetic field on holographic superconductors in
  presence of nonlinear corrections.
\newblock {\em Phys. Rev. D}, 86:106009, Nov 2012.

\bibitem{Zhao201398}
Zixu Zhao, Qiyuan Pan, Songbai Chen, and Jiliang Jing.
\newblock Notes on holographic superconductor models with the nonlinear
  electrodynamics.
\newblock {\em Nuclear Physics B}, 871(1):98 -- 110, 2013.

\bibitem{yao2013sv}
Weiping Yao and Jiliang Jing.
\newblock Analytical study on holographic superconductors for born-infeld
  electrodynamics in gauss-bonnet gravity with backreactions.
\newblock {\em Journal of High Energy Physics}, 2013(5), 2013.

\bibitem{PhysRevD.15.2752}
G.~W. Gibbons and S.~W. Hawking.
\newblock Action integrals and partition functions in quantum gravity.
\newblock {\em Phys. Rev. D}, 15:2752--2756, May 1977.

\bibitem{Balasubramanian:1999re}
Vijay Balasubramanian and Per Kraus.
\newblock {A Stress tensor for Anti-de Sitter gravity}.
\newblock {\em Commun.Math.Phys.}, 208:413--428, 1999.

\bibitem{Aprile:2010yb}
Francesco Aprile, Sebastian Franco, Diego Rodriguez-Gomez, and Jorge~G. Russo.
\newblock {Phenomenological Models of Holographic Superconductors and Hall
  currents}.
\newblock {\em JHEP}, 1005:102, 2010.

\bibitem{PhysRevB.29.1493}
T.~H. Lin, X.~Y. Shao, M.~K. Wu, P.~H. Hor, X.~C. Jin, C.~W. Chu, N.~Evans, and
  R.~Bayuzick.
\newblock Observation of a reentrant superconducting resistive transition in
  granular ba${\mathrm{pb}}_{0.75}$${\mathrm{bi}}_{0.25}$${\mathrm{o}}_{3}$
  superconductor.
\newblock {\em Phys. Rev. B}, 29:1493--1496, Feb 1984.

\bibitem{PhysRevB.51.3134}
Y.~Zhao, G.~D. Gu, G.~J. Russell, N.~Nakamura, S.~Tajima, J.~G. Wen, K.~Uehara,
  and N.~Koshizuka.
\newblock Normal-state reentrant behavior in superconducting
  ${\mathrm{bi}}_{2}$${\mathrm{sr}}_{2}$${\mathrm{cacu}}_{2}$${\mathrm{o}}_{8}$/${\mathrm{bi}}_{2}$${\mathrm{sr}}_{2}$${\mathrm{ca}}_{2}$${\mathrm{cu}}_{3}$${\mathrm{o}}_{10}$
  intergrowth single crystals.
\newblock {\em Phys. Rev. B}, 51:3134--3139, Feb 1995.

\bibitem{Horowitz:2012gst}
GaryT. Horowitz, JorgeE. Santos, and David Tong.
\newblock Optical conductivity with holographic lattices.
\newblock {\em Journal of High Energy Physics}, 2012(7), 2012.

\bibitem{Horowitz:2012gs}
Gary~T. Horowitz, Jorge~E. Santos, and David Tong.
\newblock {Further Evidence for Lattice-Induced Scaling}.
\newblock {\em JHEP}, 1211:102, 2012.

\bibitem{Nishioka:2009zj}
Tatsuma Nishioka, Shinsei Ryu, and Tadashi Takayanagi.
\newblock {Holographic Superconductor/Insulator Transition at Zero
  Temperature}.
\newblock {\em JHEP}, 1003:131, 2010.

\bibitem{Bai:2012cx}
Nan Bai, Yi-Hong Gao, Bu-Guan Qi, and Xiao-Bao Xu.
\newblock {Holographic insulator/superconductor phase transition in Born-Infeld
  electrodynamics}.
\newblock 2012.

\bibitem{Chaturvedi:2014dga}
Pankaj Chaturvedi and Pallab Basu.
\newblock {Holographic quantum phase transitions and interacting bulk scalars}.
\newblock {\em Phys.Lett.}, B739:162--166, 2014.

\bibitem{Das:2011nk}
Sumit~R. Das.
\newblock {Holographic Quantum Quench}.
\newblock {\em J.Phys.Conf.Ser.}, 343:012027, 2012.

\bibitem{Basu:2011ft}
Pallab Basu and Sumit~R. Das.
\newblock {Quantum Quench across a Holographic Critical Point}.
\newblock {\em JHEP}, 1201:103, 2012.

\bibitem{Basu:2012gg}
Pallab Basu, Diptarka Das, Sumit~R. Das, and Tatsuma Nishioka.
\newblock {Quantum Quench Across a Zero Temperature Holographic Superfluid
  Transition}.
\newblock {\em JHEP}, 1303:146, 2013.

\end{thebibliography}

\end{document}